\documentclass[aps,twocolumn,pre]{revtex4-1}

\usepackage{graphicx}
\usepackage{amsmath,mathtools}
\usepackage{color}

\usepackage[hidelinks]{hyperref}

\begin{document}

\title{Work flux and efficiency at maximum power of a triply squeezed engine}

\author{Manash Jyoti Sarmah}

\author{Himangshu Prabal Goswami}
\email{hpg@gauhati.ac.in}
\affiliation{Department of Chemistry, Gauhati University, Jalukbari, Guwahati-781014, Assam, India}
\date{\today}

\begin{abstract} 
We explore the effects of quantum mechanical squeezing on the nonequilibrium thermodynamics of a coherent heat engine with squeezed reservoirs coupled to a squeezed cavity.  We observe that the standard known phenomenon of flux- optimization beyond the classical limit with respect to quantum coherence is destroyed in presence of squeezing. Under extreme nonequilibrium conditions, the flux is rendered independent of squeezing. The efficiency at maximum power (EMP) obtained by optimizing the cavity's squeezing parameter is greater than what was predicted by Curzon and Ahlborn even in the absence of reservoir squeezing. The EMP with respect to the either of reservoirs' squeezing parameters is surprisingly equal and linear in $\eta_C$ with a slope unequal to the universally accepted slope, $1/2$.  The slope is found to be proportional to the dissipation into the cavity mode and an intercept equal to a specific numerical value of the engine's efficiency. 
\end{abstract}

\maketitle

\section{Introduction}
 One or more quantum systems that operate between two separate reservoirs make up a Quantum Heat  Engine (QHE). QHEs have the primary function of converting heat into work \cite{quan2007quantum,kosloff2014quantum,campisi2015nonequilibrium,PhysRevLett.2.262,science.342.713,PhysRevLett.122.110601,PNAS.108.15097}. Apart from traditional thermal reservoirs, the use of non-thermal baths, which are constructed reservoirs with correlated characteristics, have provided a thorough setting for examining the relationship between quantum effects and thermodynamic quantities \cite{doi:10.1126/science.1078955,PhysRevE.86.051105,PhysRevE.93.052120,PhysRevE.100.052126,PhysRevE.98.042123}.
Squeezed states or non-canonical initial states  \cite{walls1983squeezed, puri1997coherent, dupays2021shortcuts} are such non-thermal baths which allow additional control over any quantum systems' dynamics garnering  tremendous interest off late in the context of open quantum systems \cite{kumar2022thermodynamics,dupays2021shortcuts,PhysRevE.100.052126}. 

Current technologies permit experimental realization of such states \cite{klaers2017squeezed} and its effects on the thermodynamics are experimentally realizable through recently designed experimental quantum heat engines (QHE)\cite{PhysRevX.7.031044,PhysRevA.100.042119,rossnagel2014nanoscale,zou2017quantum,melo2022experimental}. Intense  efforts have been made to interrogate QHEs on the role of coherence, correlations or entanglement on the underlying dynamics \cite{Niedenzu_2016,Lostaglio_2015,PhysRevX.5.021001,Korzekwa_2016}.
 It has already been demonstrated that certain quantum resources can be exploited to bend the limits of classical thermodynamics \cite{Abah_2014,PhysRevLett.112.030602, kumar2022thermodynamics}. Coherence enhanced power and efficiency  and optimization of the flux via quantum coherences in QHEs are well studied and established phenomena \cite{PNAS.108.15097,um2022coherence,PhysRevA.88.013842,PhysRevA.86.043843,latune2021roles}. 
 Squeezed thermal baths too have proven crucial, especially in the light of a proof-of-concept experiment based on a nanobeam heat engine\cite{PhysRevX.7.031044}. Efficiency greater than that of Carnot has also been predicted \cite{rossnagel2014nanoscale}.

On the theoretical front, quantum thermodynamic analysis of QHEs s are performed by combining principles from quantum optics and nonequilibrium statistical mechanics \cite{PhysRevE.86.051105,Manzano_2016,https://doi.org/10.48550/arxiv.1706.06206,PhysRevE.91.062137}. In quantum optics, squeezing \cite{chen2006quantum,article} generally leads to less observation of  quantum noise than thermal states \cite{doi:10.1142/S021797929100033X}. Squeezing alters the entropy flow associated with the heat exchanged with the system and introduces an additional term proportional to the second-order coherences which determines the asymmetry in the second-order moments of the mode quadratures, which takes into account both the relative variance shape and the relative optical phase space displacements\cite{Manzano_2016}. This manifests in an increased efficiency, even surpassing the Carnot bound \cite{PhysRevE.93.052120,Niedenzu_2016,PhysRevB.96.104304,PhysRevX.7.031044,PhysRevE.95.032139}. To account for a realistic performance of such QHEs, usually a finite time assessment is performed by evaluating the efficiency at maximum power (EMP),  originally introduced in a classical context \cite{curzon1975efficiency}. From a nonequilibrium quantum statistical point of view, the near equilibrium EMP is universally accepted to be $\eta_C/2$ \cite{PhysRevLett.95.190602}, with $\eta_C$ being the standard Carnot efficiency of a classical heat engine. Recently, this robust expression has been showed to be invalid if the engine is locally optimized \cite{PhysRevE.98.052137}. The EMP has been shown to be modified into several forms as one keeps changing or introducing or optimizing additional system parameters \cite{PhysRevB.96.104304,PhysRevE.103.052125}.  One particularly interesting form of the EMP has been predicted recently which holds in the presence of squeezed reservoirs, being equal to $\eta_{m}^2/(\eta_{m}-(1-\eta_{m})\ln(1-\eta_{m}))$. Here, $\eta_{m}$ being a squeezing-dependent effective Carnot efficiency \cite{PhysRevE.100.052126}. However, the validity of such robust thermodynamic expressions remains questionable when engines operate in presence of both quantum coherences and quantum squeezing since the general framework on which such studies were based didn't take such effects into account. The current work is motivated on this latter aspect.

In this work, we address  how the thermodynamics of a QHE coupled to squeezed cavity respond to reservoir squeezing in presence of coherences using a quantum master equation technique. Such a technique is standard and  has already been used in nonequilibrium quantum transport studies with squeezed reservoirs \cite{abebe2021interaction,li2017production,sarmah2022nonequilibrium}. Unsqueezed dynamics of the engine that we cosider has also been well studied \cite{PNAS.108.15097,UHeplQHE,PhysRevA.86.043843}.  In Sec.(\ref{model}), we introduce our triple squeezed QHE model and its dynamics. In Sec.(\ref{thermodynamics}), we explore the effects of squeezing on the flux into the cavity mode, which we call the work-flux. In Sec.(\ref{sec-EMP}), we evaluate the EMP with respect to three squeezing parameters and a system parameter after which we conclude.


\begin{figure}[b!]
\centering
\includegraphics[width=8cm]{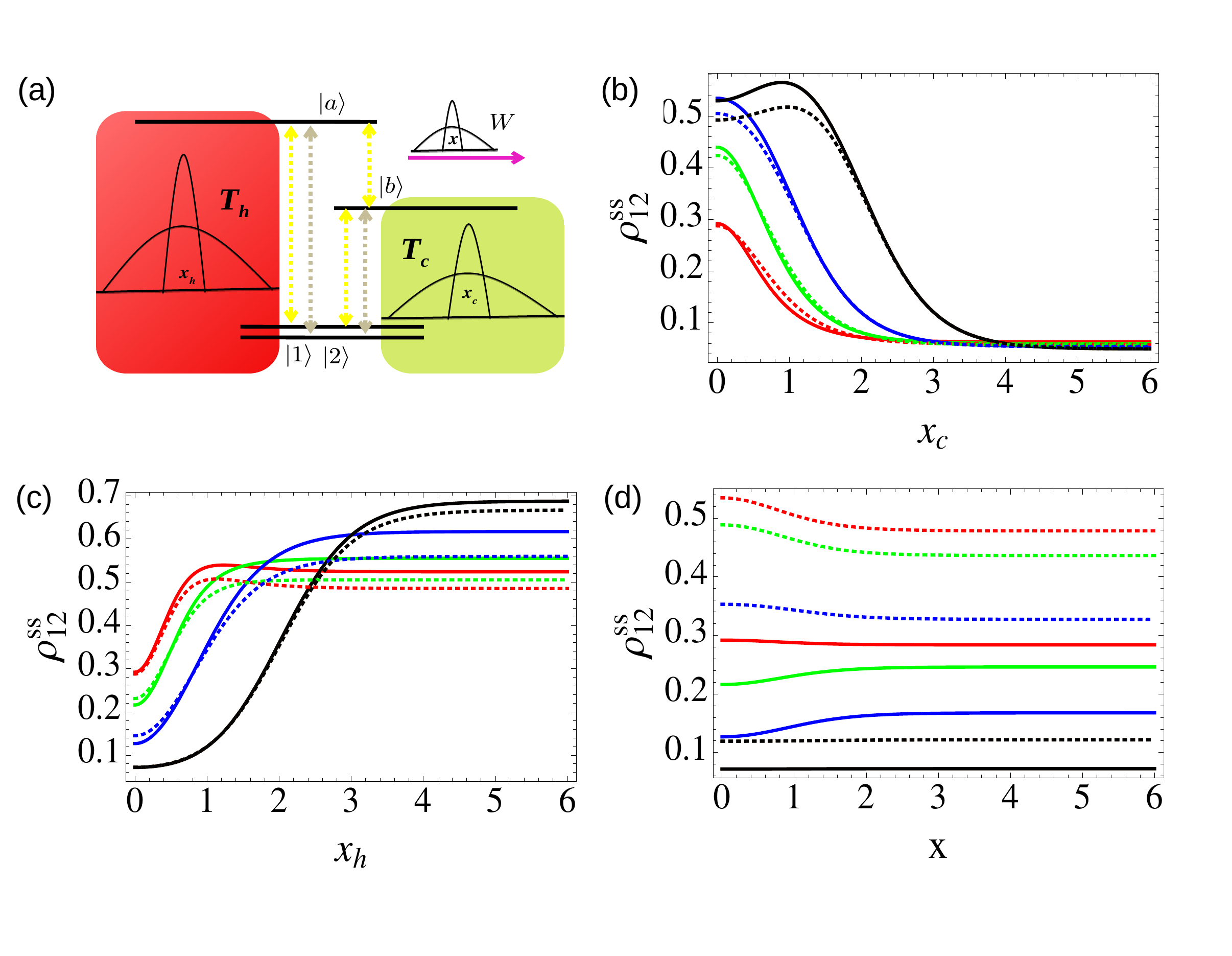}
\caption{(Color online) a) Level scheme of the model quantum heat
engine.  A pair of degenerate levels $|1\rangle$ , $|2\rangle$ is resonantly
coupled
to two excited levels $|a\rangle$ and $|b\rangle$ by two thermally populated
squeezed field modes with hot (T$_h$) and cold ($T_c$) temperatures. Levels
$|a\rangle$ and $|b\rangle$ are coupled through a squeezed cavity mode of
frequency $\nu_\ell$ . Emission of photons into this squeezed cavity is the work
done by the QHE. The engine parameters are fixed through out the manuscript at $E_1=E_2=0.1, E_b=0.4, E_a = 1.5, g = 1, r = 0.7$ and $\tau = 0.5$ in the unit of $k_B\to 1$ and $\hbar\to 1$. b) The solid (dotted) curves represent the steadystate coherence, $\rho_{12}^{ss}$ (solved by setting the RHS of Eq.(\ref{p12-eq})=0) as a function of the b) cold bath squeezing parameter $x_c$ evaluated at different values of $x_h =0,0.5,1,2$, bottom to top with $ x=1$ ($x=0$), c) hot squeezing parameter, $x_h$ with $x_c=0,0.5,1,2$, bottom to top and $x=1$ ($x=0$), d)  cavity squeezing,$x$ with the solid curves (bottom to top) evaluated at $x_h =0,x_c=0,0.5,1,2$. The dotted ones represent $x_c =0,x_h=0,0.5,1,2$.  
}
\label{c7-schematic}
\end{figure}
\section{Squeezed Engine Dynamics}
\label{model}
The QHE model consists of four quantum levels coupled asymmetrically to 
two squeezed baths with the upper two levels coupled to a squeezed unimodal cavity as shown schematically in Fig.(\ref{c7-schematic}a). Experimentally, similar QHEs have been realized in cold Rb and Cs atoms using magneto optical traps \cite{zou2017quantum,bouton2021quantum}. The squeezed density matrices of the QHE can be written as\cite{li2017production,PhysRevA.105.062219},
\begin{align}
    \bar\rho_\ell^{}&=\frac{1}{Z_\ell} \exp\{-\beta_\ell^{} \hat{S}_\ell\hat H_\ell\hat S^\dag_\ell\},\\
    \bar\rho_\nu^{}&=\frac{1}{Z_\nu} \exp\{-\beta_\nu^{} \hat{S}_\nu\hat H_\nu\hat S^\dag_\nu\},\nu=h,c,
\end{align}
with $\beta_z = (k_B T_z)^{-1}, z=\ell,h,c$ being the inverse temperatures of the cavity, hot and cold reservoirs respectively. $\hat S (\hat S_{\nu})$ is the squeezing operator on the squeezed cavity's mode (reservoirs' modes) given by :
\begin{align}
    \hat S_\ell &=e^{\frac{1}{2}(x \hat a_{\ell}^{\dag2}-h.c)},\\
    \hat S_\nu &=\displaystyle \prod_{k} e^{\frac{1}{2}(\lambda_{k\nu}^* \hat a_{k\nu}^{\dag2}-h.c)},\\
    \lambda_{k\nu} &= x_{k\nu} e^{i\theta_{k\nu}} , x_{k\nu}>0. 
\end{align}
$\theta_{k\nu}$ and $x_{k\nu}$ are the squeezing parameters of the reservoirs  and
 $x$ is the squeezing parameter \cite{dodonov2002nonclassical,li2017production,PhysRevA.105.062219,sarmah2022nonequilibrium}. $\hat H_\ell= \epsilon_\ell\hat{a}_{\ell}^{\dag}\hat{a}_{\ell}$ is the Hamiltonian for the cavity mode and $\hat H_\nu=\sum_{k}\epsilon_{k\nu}\hat{a}_{k\nu}^{\dag}\hat{a} 
        _{k\nu}$ is the Hamiltonian for the $\nu$-th reservoir. The total Hamiltonian of the four level QHE is $\hat{H}_{T}\,=\,\sum_{\nu\,=\,1,2,a,b}
        E_{\nu}|\nu \rangle\langle \nu |+\hat H_\ell+\hat H_\nu+\hat{V}_{sb}+\hat V_{sc}$, with the system-reservoir and system-cavity coupling Hamiltonians given by,
\begin{align}
\hat V_{sb}&=\sum_{k\,\in\,h.c}\sum_{i\,=\,1,2}\sum_{x\,=\,a,b}r_{ik}\hat{a}_{k}|x\rangle\langle i|+h.c\\
\hat V_{sc}&=g\hat{a}_{\ell}^\dag|b\rangle\langle a|+h.c.
\end{align}
 $\epsilon_{k}$, $\epsilon_{\ell}$ and $E_{\nu}$ denote the energy of the $k$th mode of the two thermal reservoirs, the unimodal cavity and system's $\nu$th energy level respectively. The system-reservoir coupling of the $i$th state with the $k$th mode of the reservoirs is denoted by
$r_{ik}$.  $\hat{a}^\dag (\hat{a})$ are the bosonic creation (annihilation) operators. 
The radiative decay originating from the transition $|a\rangle\to|b\rangle$ is the work done by the engine.  Unsqueezed version of such a QHE has been thoroughly studied using a Markovian quantum master equation  \citep{PhysRevA.86.043843,PNAS.108.15097,PhysRevA.88.013842,UHeplQHE,PhysRevE.99.022104}. Following such a  standard procedure to derive of a quantum master equation \cite{UHeplQHE,li2017production} for the matrix elements of the reduced density matrix $\rho$ (supplementary information)  has four populations, $\rho_{ii}, i = 1, 2, a ,b$ coupled to the real part of a coherence term, $\rho_{12}$. The coherence $\rho_{12}$ between states $|1\rangle$ and $|2\rangle $ arise due 
to interactions with the hot and the cold baths. This thermally induced coherence couples
to populations due to transition involving the states $|1\rangle$ 
and $|2\rangle$.
Under the symmetric coupling regime, we can now write down five coupled first order differential equations describing the time-evolution of the four populations and the coherence (under symmetric coupling, $r$), given by

\begin{eqnarray}
\dot{\rho}_{12}&=&\frac{-ry}{2} \rho_{11} - \frac{ry}{2}\rho_{22}+r
p_h\tilde{N}_h\rho_{aa}+rp_c\tilde{N}_c \rho_{bb}\nonumber\\&-&r (n+ \tau)
\rho_{12}\label{r12}
\label{p12-eq}\\
\dot{\rho}_{ii}&=&-rn \rho_{ii} + r\tilde{N}_h\rho_{aa} +  \tilde{N}_c
\rho_{bb}- ry\rho_{12}, i = 1,2\label{r11}
\label{pii-eq}\\
\dot{\rho}_{bb}&=&rN_c \rho_{11} + rN_c \rho_{22} + 
 g^2 \tilde{N}_\ell \rho_{aa}\nonumber\\&-& (g^2 N_\ell + 2r\tilde{N}_c)
\rho_{bb}+2rp_c N_c\rho_{12}\label{rbb}\\
\dot{\rho}_{aa}&=&r N_h \rho_{11} + r N_h \rho_{22} - (g^2 \tilde{N}_\ell + 2 r
\tilde{N}_h) \rho_{aa}\nonumber\\
&+& g^2 N_\ell \rho_{bb}+2 r p_h N_h \rho_{12}\label{raa}
\end{eqnarray}
with, $\sum_i\rho_{ii}=1, i=1,2,a,b$ and $n=N_c+N_h$, $y=N_cp_c+N_hp_h$, with the reorganized occupation factors given by
\begin{eqnarray}
N_z&=&\cosh(2x_z)(n_{z}+\frac{1}{2})-\frac{1}{2},z=h,c,\\
N_\ell&=&\cosh(2x)(n_{\ell}+\frac{1}{2})-\frac{1}{2}.
\end{eqnarray}
Here, $n_c,n_h$ and$  n_l$ are the Bose-Einstein distributions for the cold reservoir, hot reservoir and the cavity respectively. These factors are now squeezing dependent via the dimensionless parameters, $x_h,x_c$ and $x$ representing the extent of squeezing in the hot, cold reservoirs and the cavity respectively. 
$p_\nu=|\cos\phi_\nu|,\nu=h,c$ are two dimensionless parameters that governs the strength of
coherences and whose values are dictated by the angles of relative orientation ($\phi_\nu$) of the $\nu-$th bath induced transition in the system \cite{PNAS.108.15097,UHeplQHE,PhysRevE.99.022104}. A phenomenological dimensionless rate $\tau$ has been  added 
to take care of the dephasing. 
Setting $\dot\rho=0$, at the steady state, we can solve for the
steady state values of $\rho_{aa}$,
$\rho_{bb}, \rho_{11},\rho_{22}$, and $\rho_{12}$ and obtain these analytically (supplementary text).

\begin{figure}
\centering
\includegraphics[width=8cm]{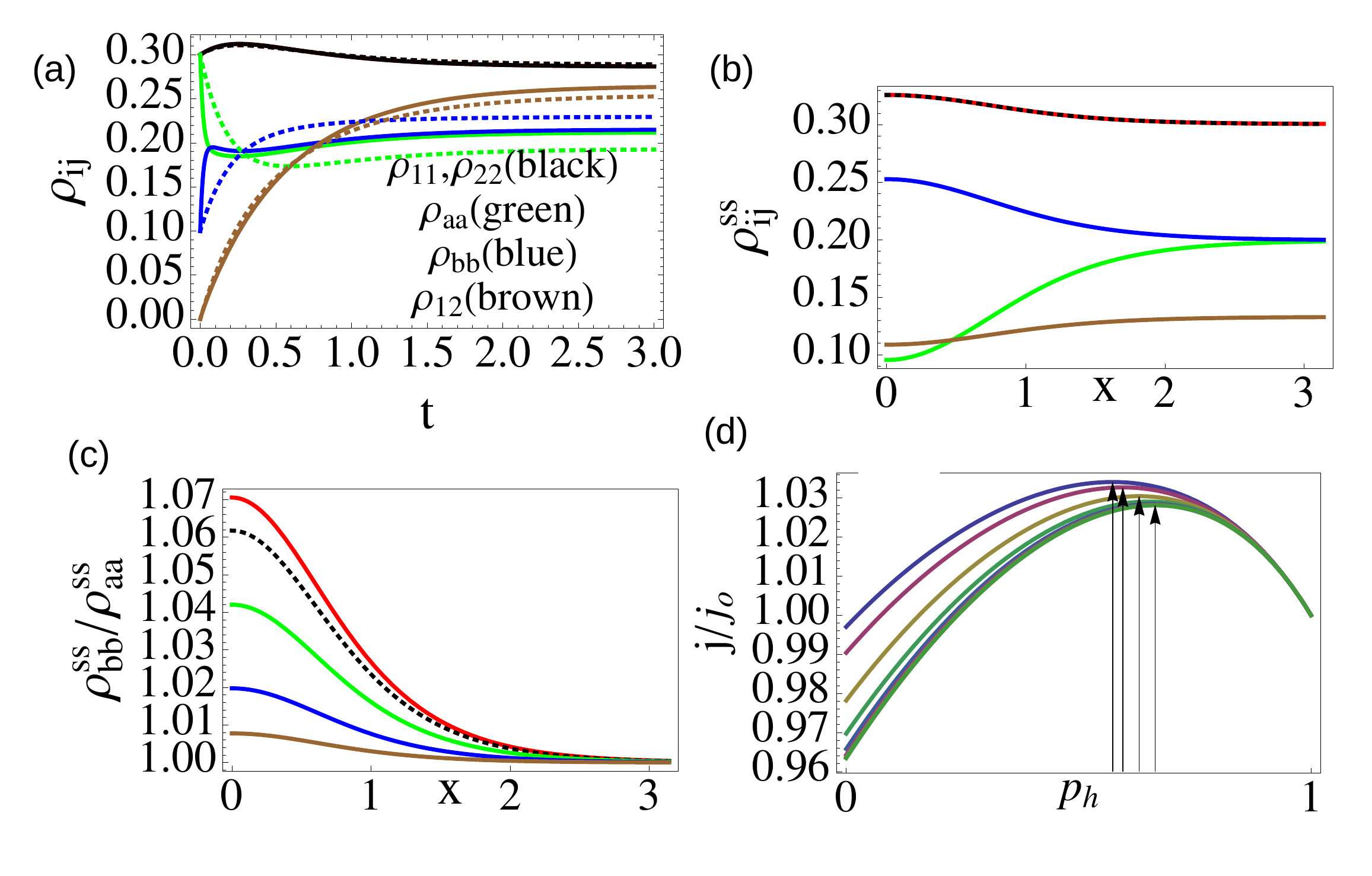}
\caption{a) The solid (dotted) curves represent time evolution of $\rho_{ij}$ with, $x=2$ (without, $x=0$) squeezing obtained by solving Eq.(\ref{p12-eq}-\ref{raa}) for $T_h=2,T_c=0.5,T_l=0.9$. b) Steady state values as a function of the squeezing parameters for the same parameters as (a) c) Ratio of the steady state values between states $|b\rangle$  and $|a\rangle$ reaching unity highlighting the equipopulated nature under high squeezing; $p_c =0.2,0.3,0.5,0.7,0.8$ from the top to the bottom curves. (d) Optimization of the flux ratio as a function of hot coherence parameter, $p_h$ for different squeezing parameters under far from equilibrium conditions and $p_c=1$ (top to bottom: $x =0, \pi/6,\pi/\pi/2,2\pi/3,5\pi/6,\pi,3\pi/2$). Other parameters are same as Fig.(\ref{c7-schematic}a).  }
\label{fig-pop}
\end{figure}

The steadystate value of the coherence term $\rho_{12}^{ss}$ as a function of the squeezing parameters, $x_h,x_c$ and $x$ are shown in Fig.(\ref{c7-schematic}b,c,d)) for different engine parameters. The different curves in  Fig.(\ref{c7-schematic}b) represent $\rho_{12}^{ss}$ evaluated for different $x_h$ and $x$ values as a function of $x_c$. The solid (dotted) lines represent $\rho_{12}^{ss}$ when $x_h\ne0 (x_h=0)$ and $x=0 (x\ne0)$. At high $x_c$ values, the coherence is reduced and saturates to a lower value in comparison to $\rho_{12}^{ss}$ values of lower $x_c$. At high $x_h$ values (black curve), $\rho_{12}^{ss}$ steadily increases and reaches a maximum value around some intermediate $x_c$ value and then sharply drops as $x_c$ keeps increasing. This behavior is however absent for lower $x_h$ values. Fig.(\ref{c7-schematic}c) represent $\rho_{12}^{ss}$ evaluated for different $x_c$ and $x$ values as a function of $x_h$. The solid (dotted) lines represent $\rho_{12}^{ss}$ when $x_c\ne0 (x_c=0)$ and $x=0 (x\ne0)$. At high $x_h$ values, the steady state values of the coherence term  increases and saturates to a higher value in comparison to coherence at lower $x_h$ values. We can rationalize that, $x_c(x_h)$ tend to reduce (increase) the steadystate values of the coherences as we keep squeezing the baths more and more. The same however cannot be said for $\rho_{12}^{ss}$ vs $x$ as seen from Fig. (\ref{c7-schematic}d). The solid (dotted) lines represent the behavior at $x_h=0(x_h\ne 0)$ for finite $x_c$ values. 

The time evolution of each of the  equations (Eq.(\ref{r12}-\ref{raa})) for various engine parameters for $x_h=x_c=0$ and $x=2$ is shown in Fig.(\ref{fig-pop}a). In Fig.(\ref{fig-pop}b), the steadystate values of the populations as a function of $x$ is shown   where solid (dotted) curves represent cavity-squeezed, $x\ne 0$ (cavity-unsqueezed, $x = 0$) evolutions. Note that under high squeezing of the cavity mode, the steady state values, $\rho_{aa}^{ss}$ and $\rho_{bb}^{ss}$ equipopulate giving,
\begin{align}
 \label{eq-aa=bb}
 \displaystyle\lim_{x\to \infty}\frac{\rho_{bb}^{ss}}{\rho_{aa}^{ss}}&=1
\end{align}
and is shown numerically in Fig.(\ref{fig-pop}c) for different values of the hot coherence parameter, $p_h$. The analytical expressions for the steadystate values are provided in the supplementary information.

\begin{figure}
\centering
\includegraphics[width=8cm]{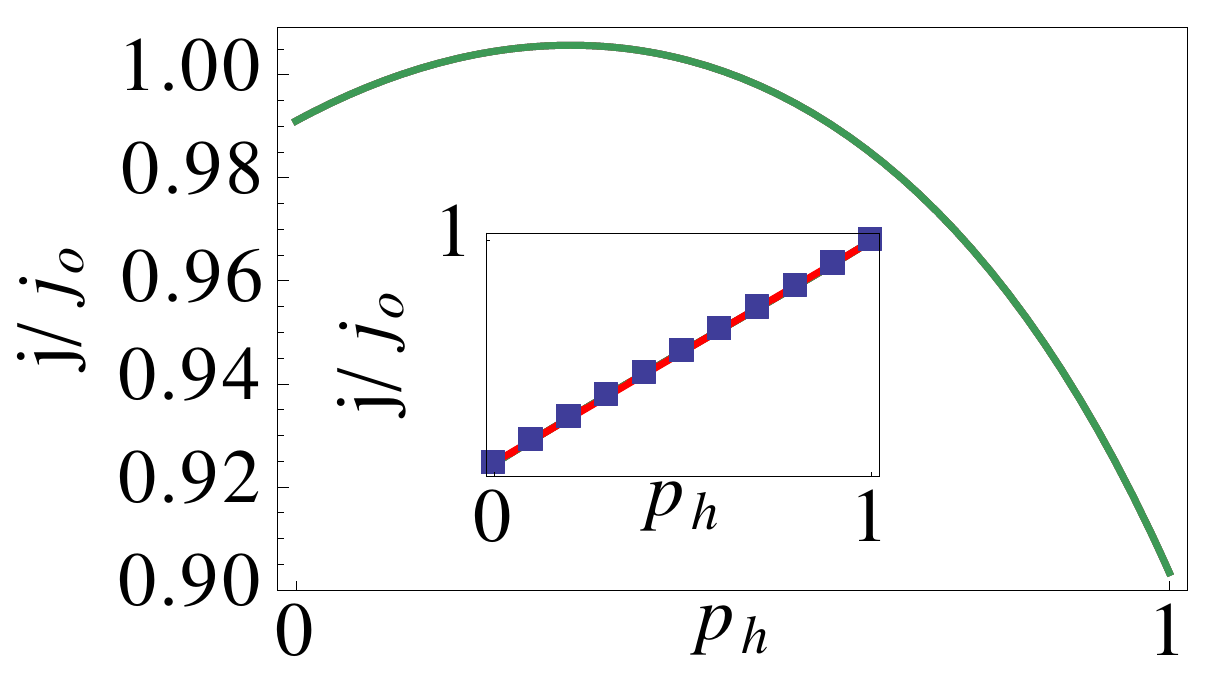}
\caption{ Failure of coherence to optimize the flux beyond classical values ($j/j_0>1$) under high squeezing as given by Eq.(\ref{eq-flux-xinf}). Inset: Linear dependence of the flux ratio on $p-j$ under high squeezing ($x\gg 0$) and  $T_l\gg 0$ given by Eq.(\ref{eq-rat}) evaluated at $p_c=1, T_c =0.5,T_h=1$. The square boxes represent linear fit.}
\label{fig-failure}
\end{figure}

\begin{figure}
\centering
\includegraphics[width=8cm]{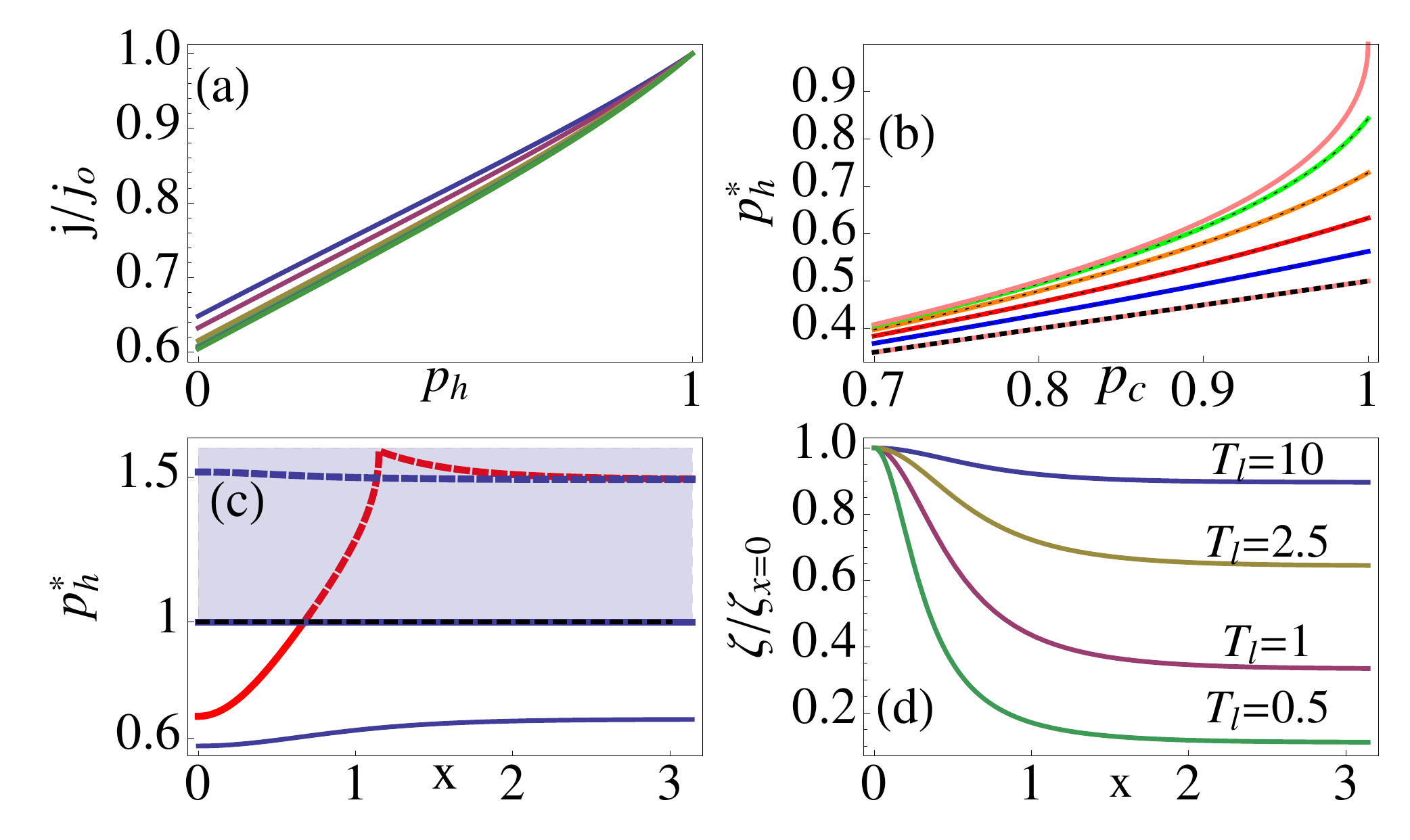}
\caption{ a) Loss of optimization of flux as a function of $p_h$ for different squeezing parameters, near equilibrium ($T_c=0.9, T_h = 1, T_l=10$). (b)  Loss of linear dependence of $p_c$ on the optimal value $p_h^*$ as given by Eq.(\ref{eq-phmax-hi}). The topmost curve represents Eq.(\ref{eq-phmax-hi-tc0}). (c) Plot showing breakdown of the coherent optimization of the flux as a function of squeezing parameter. The shaded region is not allowed since the maximum possible value of $p_h^*$ is unity. Under far from equilibrium condition $p_h^*$ exists which saturates (bottom curve) at higher values of $x$ given by Eq.(\ref{eq-phmax-hi}). The top curve shows the behavior of $p_h^*$ near equilibrium  which is nonexistent after a certain squeezing value.  (d) (d) Lowering of thermodynamic affinity as a function of squeezing evaluated at $T_c=0.1, T_h = 2$ and $x_c=x_h=0$. }
\label{fig-ph}
\end{figure}

\section{Work Flux}
\label{thermodynamics}
We interprete the emission of photons into the squeezed cavity as the work done by the engine. This photon exchange process between the levels $|a\rangle, |b\rangle$ with the squeezed cavity is quantified by the rate of photon exchange with the cavity which we refer to as the work flux,
$j=\frac{d}{dt}\langle a_\ell^\dag a_\ell\rangle$, where the trace is with respect to the squeezed cavity density matrix. Following a standard procedure to second order in
the coupling as developed in\cite{PhysRevA.88.013842,UHeplQHE} we get, $j=g^2(\tilde N_\ell\rho_{aa}^{ss}-N_\ell\rho_{bb}^{ss})$. We can substitute the values of the steadystate populations to obtain an analytical expression for the flux (supplementary information).
When, the hot and the cold coherence parameters  individually go to zero ($p_c=p_h=0$), the coherence vanishes ($\rho_{12}^{ss}$=0) and we obtain a coherence -unaffected value of the flux, which we denote as $j_o$. Note that, $j_o$ depends on the squeezing parameters $x,x_h$ and $x_c$. In 
the absence of squeezing ($x_h=x_c=x=0$), $j_o$ shall be denoted by $j_o^0$,
 which we refer to as  the classical value of the flux. There are no effects of coherence or squeezing on $j_o^0$. It is a well known phenomena that, in absence of squeezing, $j>j_o$ can be achieved as a function of coherence parameter, $p_h$ \cite{PNAS.108.15097,um2022coherence}. We plot the ratio $j/j_o$ in shown in Fig.(\ref{fig-pop}d) for different squeezing values of the cavity for $x_c=x_h=0$.  As the cavity squeezing parameter is increased the optimal value of the flux gradually decreases and the $p_h$ value that optimizes the ratio (denoted as $p_h^*$) shifts towards larger $p_h$ values.   
 We now attempt to explore the dependence of the flux in presence of squeezing on the coherences in detail. Since the analytical expressions of $j$ and $j_o^0$ are too lengthy we focus on some limiting cases.
 
 Under high cavity squeezing, ($x\to \infty$), we obtain $\rho_{aa}^{aa}=\rho_{bb}^{ss}$ as seen from Eq.(\ref{eq-aa=bb}).  The expression for the flux  in this case is simply given by,
 \begin{align}
  \label{eq-jxinf}
  \lim_{x\to\infty}j=g^2(\lim_{x\to\infty}\rho_{aa}^{ss}),
 \end{align}
which
under the condition $p_c=0,p_h=0$ in Eq.(\ref{eq-jxinf}) is,
\begin{align}
\label{eq-joxinf}
 \lim_{x\to\infty}j_o=\frac{r (N_h- N_c)}{2 ( n+1)}.
\end{align}
 Eq. (\ref{eq-jxinf}), with $p_c=1$  can be expressed as,
\begin{align}
 \label{eq-flux-xinf}
 \lim_{x\to\infty} j|_{p_c= 1}&=\frac{r ( N_h-N_c) \left(N_h \left(1-p_h^2\right)+t\right)}{(1-p_h) f_n+2 \tau ( n+1)}
\end{align}
with $f_n = 4  N_c
   N_h+ n (2 N_h (p_h+1)+p_h+2)$. The RHS of Eq.(\ref{eq-joxinf}) is always greater than RHS of Eq.(\ref{eq-flux-xinf}) as seen from the numerical result in Fig (\ref{fig-failure}). The physical interpretation is that the coherences are no longer able to increase the flux beyond the non coherence values. Under this condition, the ratio is bounded below unity as seen in Fig.(\ref{fig-failure}). We can analytically prove this by invoking a few conditions. In Eq.(\ref{eq-joxinf}) and (\ref{eq-flux-xinf}), if  $\tau = 0$ and $ N_h = z N_c$, $z$ being a positive integer), the ratio between the two fluxes becomes,
\begin{align}
\label{eq-rat}
 \displaystyle\frac{\displaystyle\lim_{x\to \infty}j|_{p_c=1}}{\displaystyle\lim_{x\to \infty}j_o}\bigg{|}_{N_h= zN_c}&=\! \frac{2z (p_h\!+\!1)(N_c z\!+N_c\!+1)}{2 N_c (p_h\!+\!1) z^2\!+\!z (4 N_c\!+p_h\!+\!2)\!+\!1}
\end{align}
which is a rational fraction of two linear terms of $p_h$. Eq.(\ref{eq-rat}) can be shown to have a linear dependence on $p_h$ for some appropriate conditions of the coefficients which is graphically shown as an inset in Fig.(\ref{fig-failure}). In Eq.(\ref{eq-rat}), for $z=1$ and $ N_h = N_c$ (no bias), we see a flux value that solely depends on only the coherence value, given by
\begin{align}
 \lim_{x\to \infty}\displaystyle\frac{j}{j_o}|_{N_h= N_c} &= \frac{2(1+p_h)}{(3+p_h)}\\
 &\le 1
\end{align}
and is linear in $p_h$ for small values as seen in the inset of Fig.(\ref{fig-failure}) and in Fig.(\ref{fig-ph}a). In Fig.(\ref{fig-ph}a), the flux ratio $j/j_o$ is plotted for different squeezing parameters. The squeezing decreases from top to bottom. For smaller $p_h$, the linearity is prominent, but for higher $p_h$ values, the linearity is gradually less apparent as the squeezing parameter increases.

It has been previously reported that $p_h^*$ increases linearly in $p_c$ under the unsqueezed case \cite{PhysRevA.88.013842}. In the current case, we observe that under an extremely biased scenario ($N_h\gg0$) and high squeezing, $x\gg0$, the linear dependence is lost as shown graphically in Fig.(\ref{fig-ph}b) and the dependence of $p_h^*$ on the cold coherence parameter, $p_c$ is given by the nonlinear function,
\begin{eqnarray}
 \label{eq-phmax-hi}
 p_h^*|=\frac{\sqrt{\left(1\!-\!p_c^2\right) \left(4 N_c^2 \left(1\!-\!p_c^2\right)\!+\!4 N_c\!+\!1\right)}\!+\!2 N_c
   \left(p_c^2\!+\!1\right)\!+\!1}{4 N_c p_c+p_c}\nonumber\\
\end{eqnarray}
which reduces to unity when $p_c=1$  as seen in the Fig. (\ref{fig-ph}b). The nonlinear dependence takes a simplistic form when $T_c\to 0$, where the above expression reduces to,
\begin{align}
\label{eq-phmax-hi-tc0}
 p_h^*|_{T_c=0}&=\frac{1-\sqrt{1-p_c^2}}{p_c}
\end{align}
which is shown as the topmost curve in Fig.(\ref{fig-ph}b). The RHS of Eq.(\ref{eq-phmax-hi}) also has a strange dependence on the cavity squeezing parameter. $p_h^*$ increases as a function of $x$ and saturates at higher $x$ values as shown in the bottom-most curve of Fig.(\ref{fig-ph}c). However under extremely biased conditions, $p_h^*$ sharply rises beyond unity and goes to the shaded region. The shaded region is not allowed as the maximum value of $p_h^*$ is unity.  Since an  analytical expression of $p_h^*$ as a function of $x$ is beyond the scope of simplistic analysis, the exact identification of this numerical fallout range is not possible. We simply speculate that such a breakdown happens when the cavity temperature $T_\ell$ is  set to be very high. Since $n_\ell$ is a function of $T_\ell$, the numerics blows when there is competition between $x$ and $T_\ell$ to dominate the behavior. The upper dashed curve in the shaded portion also corresponds to an unrealistic  $p_h^*$ evaluated at a high cavity temperature. In Fig.(\ref{fig-ph}d), we plot the thermodynamic force as a function of squeezing. The force can be identified from the analytical expression of the flux (supplementary text) and is given by,
\begin{align}
 \label{eq-aff}
 \zeta=\frac{\tilde{N}_c\tilde{N_\ell}N_h}{N_c\tilde{N}_hN_\ell}.
\end{align}
When $\zeta>(<)1, j>(<)1$. 
In Fig.(\ref{fig-ph}d), we plot the ratio between the thermodynamic forces in presence and absence of squeezing for different cavity temperatures. As squeezing increases, the ratio decreases for a fixed set of engine parameters and then saturates. This leads to lower magnitude of the flux in comparison to the unsqueezed case and is more prominent when the cavity temperature is low.

\begin{figure}
\centering
\includegraphics[width=8cm]{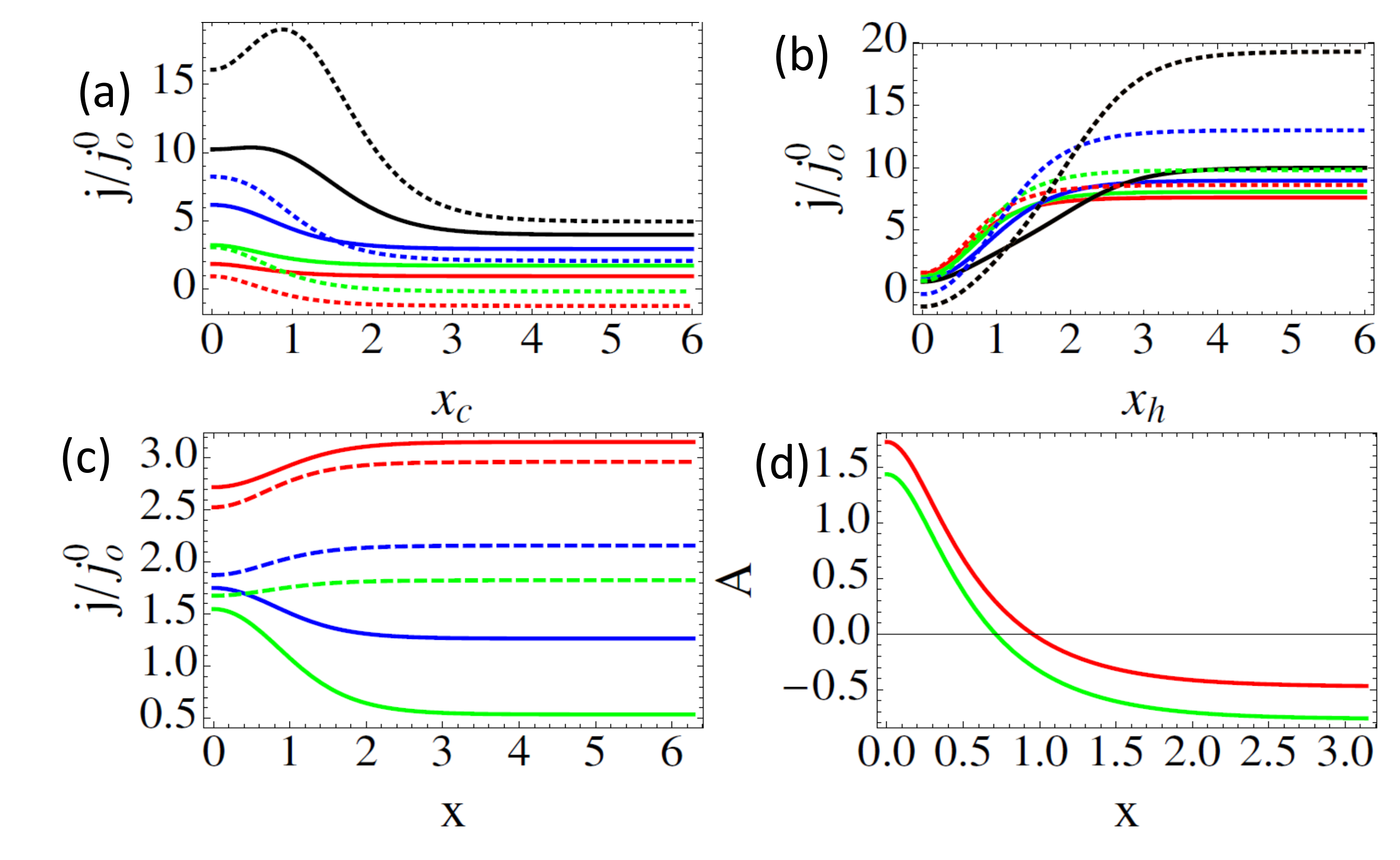}
\caption{Giant increase of the total flux (in presence of squeezing as well as coherence) in comparison to the classical case. The solid (dotted) lines represent the ratio between the total flux $j$ and the classical flux $j_o^0$   as a function of a) $x_c$ evaluated at $x=0(1), x_h = 0,0.5,1,2$, b) $x_h$ evaluated at $x=0(1), x_c = 0,0.5,1,2$. c) Solid (dotted) curves indicate the total flux ratio as a function of cavity squeezing $x$ evaluated at $T_c=0.5(0.1)$ with $\{x_h,x_c\}=\{0.5,0.1\},\{0.1,0.5\}, \{0,0\}$ (top to bottom). (d) Change in the sign of the thermodynamic affinity, $A=\log\zeta$ as function of cavity squeezing parameter evaluated at$\{x_h,x_c\}=\{1,0.1\}$ (upper curve) and $\{0.1,1\}$ (lower curve). The sign change happens at $x^*$ given by Eq.(\ref{eq-xstar}).}
\label{fig-jxch}
\end{figure}
 In Fig.(\ref{fig-jxch}a,b and c), we plot the ratio between the total flux $j$ and the classical flux $j_o^0$ as a function of $x_c, x_h$ and $x$ respectively for the same parameters as Fig.(\ref{fig-pop}). As a function of both the baths' squeezing parameters, the increase of the total flux is quite large in comparison to the classical case.  All of the curves show saturation behavior. Particularly interesting is the ratio's dependence on $x_h$ where the saturation value of the ratio is always greater than unity.

 We now focus on an extreme biased case ($T_h\gg T_c$, a limit which we invoke by taking $T_h\to\infty$ and $T_c\to 0$), a scenario when the temperature gradient is very high. This case is different from a standard extreme nonequilibrium case where the thermodynamic force must be very high ($\zeta\gg 0$).
 Under the high temperature gradient scenario, the steadystate populations of the upper two states are given by,
 \begin{align}
 \label{eq-aa-highbias}
  \lim_{T_h\gg T_c}\rho_{aa}^{ss} &=\frac{\left(p_h^2+1\right) \left(g^2  N_\ell+2 r\right)}{g^2 \left(4  N_\ell+p_h^2+1\right)-2
   \left(p_h^2-3\right) r}\\
   \label{eq-bb-highbias}
  \lim_{T_h\gg T_c}\rho_{bb}^{ss} &=\frac{g^2 \tilde N_\ell \left(p_h^2+1\right)}{g^2 \left(4  N_\ell+p_h^2+1\right)-2 \left(p_h^2-3\right) r},
 \end{align}
which no longer depends on the squeezing parameters of the two baths. Using these above values the flux can be recast as,
 \begin{align}
  \label{eq-biasedflux}
  \lim_{T_h\gg T_c}j&=\frac{2g^2r\tilde N_\ell(1+p_h^2)}{g^2(1+4N_l+p_h^2)-2r(p_h^2-3)}
  \end{align}
while the coherence-unaffected value of the flux is simply,
 \begin{align}
  \label{eq-biasedflux0}
  \lim_{T_h\gg T_c}j_o&=\frac{2 g^2 \tilde N_\ell r}{g^2 (1+4 N_\ell)+6 r}
 \end{align}
It is interesting to note that, in this highly biased scenario, the flux expression (RHS of Eq.(\ref{eq-biasedflux})) doesn't depend on the cold coherence parameter any more. In the above  two expressions, if we invoke the high squeezing scenario ($x\to \infty$), we can write down the ratio between the two fluxes as,
\begin{align}
 \label{eq-biasedfluxxinf}
  \displaystyle\lim_{x\to\infty}\frac{\displaystyle\lim_{T_h\gg T_c}j}{\displaystyle \lim_{T_h\gg T_c}j_o}&=(1+p_h^2)
\end{align}
Note that, the above expression is bound, $1\le 1+p_h^2\le 2$. In this limit with $p_h =1 (p_c\ne 1)$, coherences can double the value of the flux from its zero coherence value. Likewise, the ratio between the flux in this limit and the classical value of the flux can be written as,
\begin{align}
 \label{eq-biasedfluxxinf0}
  \displaystyle\lim_{x\to\infty}\frac{\displaystyle\lim_{T_h\gg T_c}j}{\displaystyle \lim_{T_h\gg T_c}j_o^{0}}&=
  (1+p_h^2)(1+\displaystyle\frac{6r-3g^2}{4g^2\tilde n_\ell})\\
  &\ge 1.
\end{align}
As long as $r>g^2/2$ and $p_c\ne p_h$, within the high bias scenario and maximal cavity-squeezing, the flux is always greater than unity in comparison to the classical case.

%

\begin{figure}
\centering
\includegraphics[width=8cm]{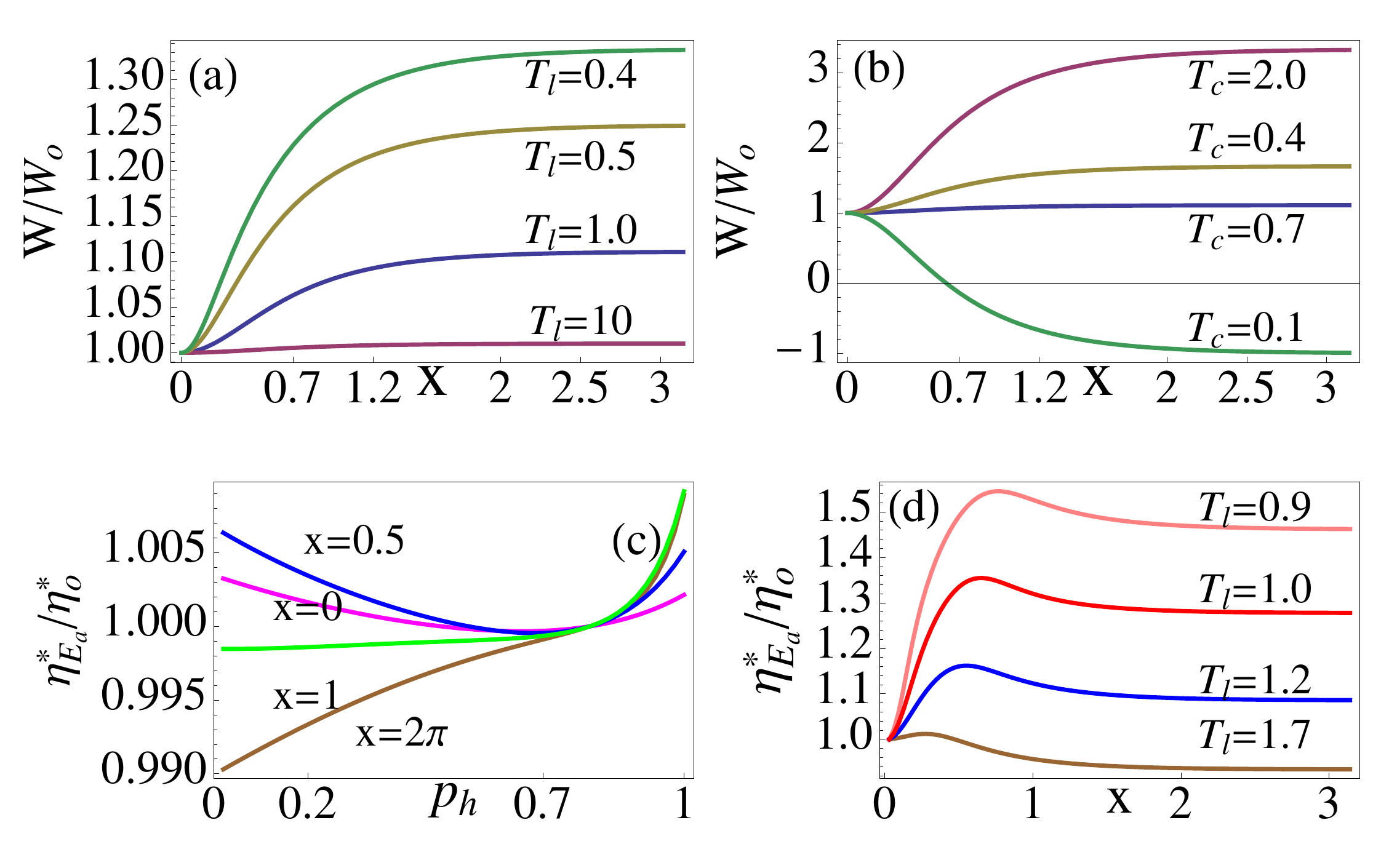}
\caption{(a) Squeezing induced increase of the work done beyond classical limits ($T_c=0.1,T_h=1$). The increase is larger when the cavity temperature is lower. (b) Negative work done as a function of squeezing for different $T_c$( $T_h=2, T_l=1$). (c) EMP with respect to $E_a$ as a function of $p_h$ for different squeezing values. (d) EMP with respect to $E_a$ for the range of squeezing at different cavity temperatures ($p_c=0.1,p_h=1$).  }
\label{fig-work}
\end{figure}
\section{Efficiency at Maximum Power}
\label{sec-EMP}
We now move to perform a thorough analysis on the efficiency  at maximum power (EMP or $\eta^*$). In a standard context, the EMP is calculated by maximizing the efficiency with respect to a system parameter. In our QHE model, the efficiency is defined as $\eta = W/Q_h$ with $Q_h^{}=(E_a-E_1)$, and the useful work done (W) is defined as,
\begin{eqnarray}
 \label{eq-useful-work2}
W = E_a-E_b-W_{diss} T_c ,
\end{eqnarray}
with $W_{diss} = k_B\mbox{ln}\frac{\tilde{N}_\ell}{N_\ell}$ is the dissipation into the cavity mode \cite{PhysRevA.88.013842,UHeplQHE}. $W$ doesn't depend on the squeezing parameters of the two squeezed reservoirs or the noise induced coherences. In Fig. (\ref{fig-work}a), we show the variation of $W/W_o$ ($W_o$ being the useful work in absence of squeezing, $x= 0$) as a function of $x$ for several values of the cavity temperature, $T_l$. As can be seen, the work done increases as $T_l$ is lowered and saturates at higher values of $x$ and  is always greater than unity as long as $T_c>T_\ell$. When $T_c<T_\ell$ (Fig.(\ref{fig-work})b), the work done is negative. In general, the work changes its sign at $x= x_{*}$, given by
\begin{align}
\label{eq-xstar}
  x_*=\frac{1}{2} \Re\left(\cosh ^{-1}\left(\frac{\tilde  N_cN_h+N_c\tilde N_h}{(2 n_\ell+1)
   (N_c-N_h)}\right)\right).
\end{align}

Although $W$ and $\eta$ are independent of coherences and the reservoir squeezing parameters, the EMP however depends on these parameters. The EMP  obtained by maximizing $P$ with respect to any system parameter puts an implicit dependence via the optimized value of the chosen parameter. We choose the three squeezing parameters $x_c,x_h$, $x$ and  $E_a$ to optimize the EMP and denote these by $\eta^*_{xc},\eta^*_{xh}, \eta^*_{x}$ and $\eta^*_{E_a}$ respectively. The squeezing unaffected values of the EMP are denoted by $\eta_o^*$. In Fig.(\ref{fig-work}c), we show the dependence of the ratio $\eta^*_{E_a}/\eta^*_o$ as a function of $p_h$ for several $x$-values evaluated at $x_c=x_h=0$ and $p_c=0.9$. The dependence of this ratio on $p_h$ is extremely nonlinear and is unity at $p_h =0.8 $ where effects of coherence vanish. At lower (higher) squeezing values, the ratio decreases (increases) to unity and then sharply increases  beyond unity as a function of $p_h$. We can theorize that,  lower $p_h$ values (under the condition $p_h<p_c$), smaller values of  cavity squeezing favor increasing the EMP beyond classical values while for larger $p_h$ ($p_h>p_c$), high squeezing favor increase of the EMP beyond classical values. In Fig.(\ref{fig-work}d), we plot the same ratio as a function of cavity squeezing parameter for different cavity temperatures, $T_l$. There is an optimization of the EMP at lower values of $x$ and the hump keeps shifting leftward to even smaller values as $T_l$ is increased and the EMP ratio keeps decreasing. From Fig.(\ref{fig-work}d), we can conclude that lower values of $T_\ell$ yield very high values of EMP with respect to $E_a$ under moderate squeezing conditions of the cavity.
\begin{figure}[t!]
\centering
\includegraphics[width=8cm]{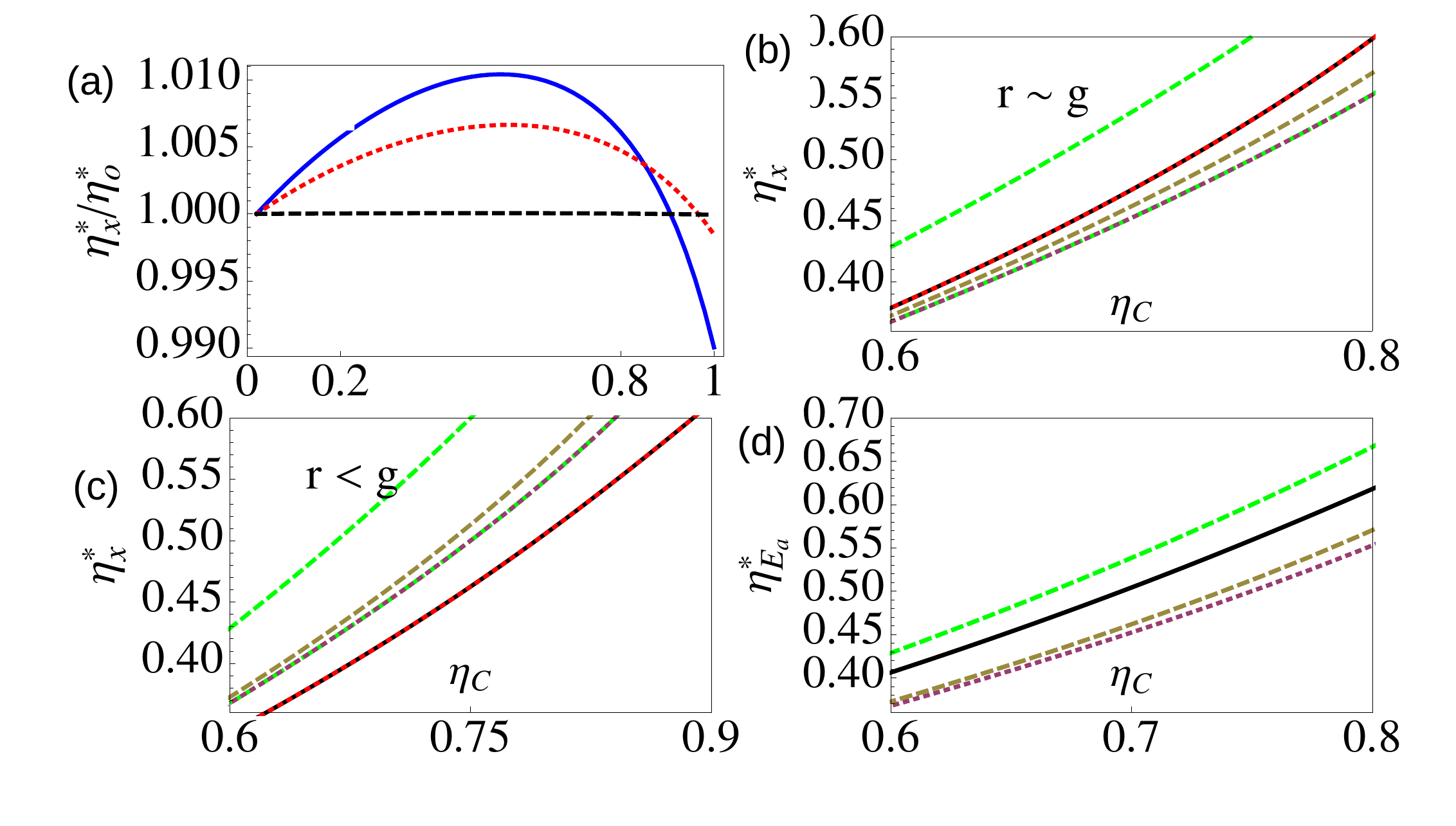}
\caption{(Color online)(a) EMP with respect to squeezing as a function of $p_h$ fr various $p_c$. In a), b) and c), the black curves (overlayed with red color) represent the evaluated EMP of our QHE. The green dashed curve is the upper bound on the EMP, $\eta^{**}$. The brown dashed line represents $\eta_{CA}$. The dotted line represent $\eta_L$. (b) and (c) EMP with respect to $x$ as a function of $\eta_C$ with $r=0.7, g = 1)$ and  $r = 0.1, g = 3$ respectively. When $r\approx g,$ $\eta_x^*> \eta_{CA}$ as seen in (b). (d) EMP with respect to $E_a$ as a function of $\eta_C$ with $r=0.7, g = 1).$ Here, $\eta_{E_a}^*> \eta_{CA}$ with $x=1(x_c=x_h=0)$.  }
\label{fig-emp}
\end{figure}
In Fig. (\ref{fig-emp}a), we plot $\eta^*_x$ as a function of $p_h$ for different combinations of $x_c$ and $x_h$ for a fixed $p_c$ value (0.5). Here, for a fixed set of engine parameters, when $x_c<x_h$ leads to a larger optimized value (around $p_h=0.5$) of the EMP with respect to $x$ (blue curve in the figure). However as $p_h$ approaches unity, there is a sharper fall in the EMP and goes below unity. For the case when $x_c=x_h$, the behavior is similar (dotted curve) but the increase is not as high as the previous case. When squeezed to the limits, $x_c\to\infty, x_h\to\infty$, the EMP with respect to $x$ no longer depends on the coherence (dashed curve). This is due to the fact that, under this scenario, the power cannot be optimized with respect to $x$ and the maximum value occurs at $x=0$.

\begin{figure}
\centering
\includegraphics[width=8cm]{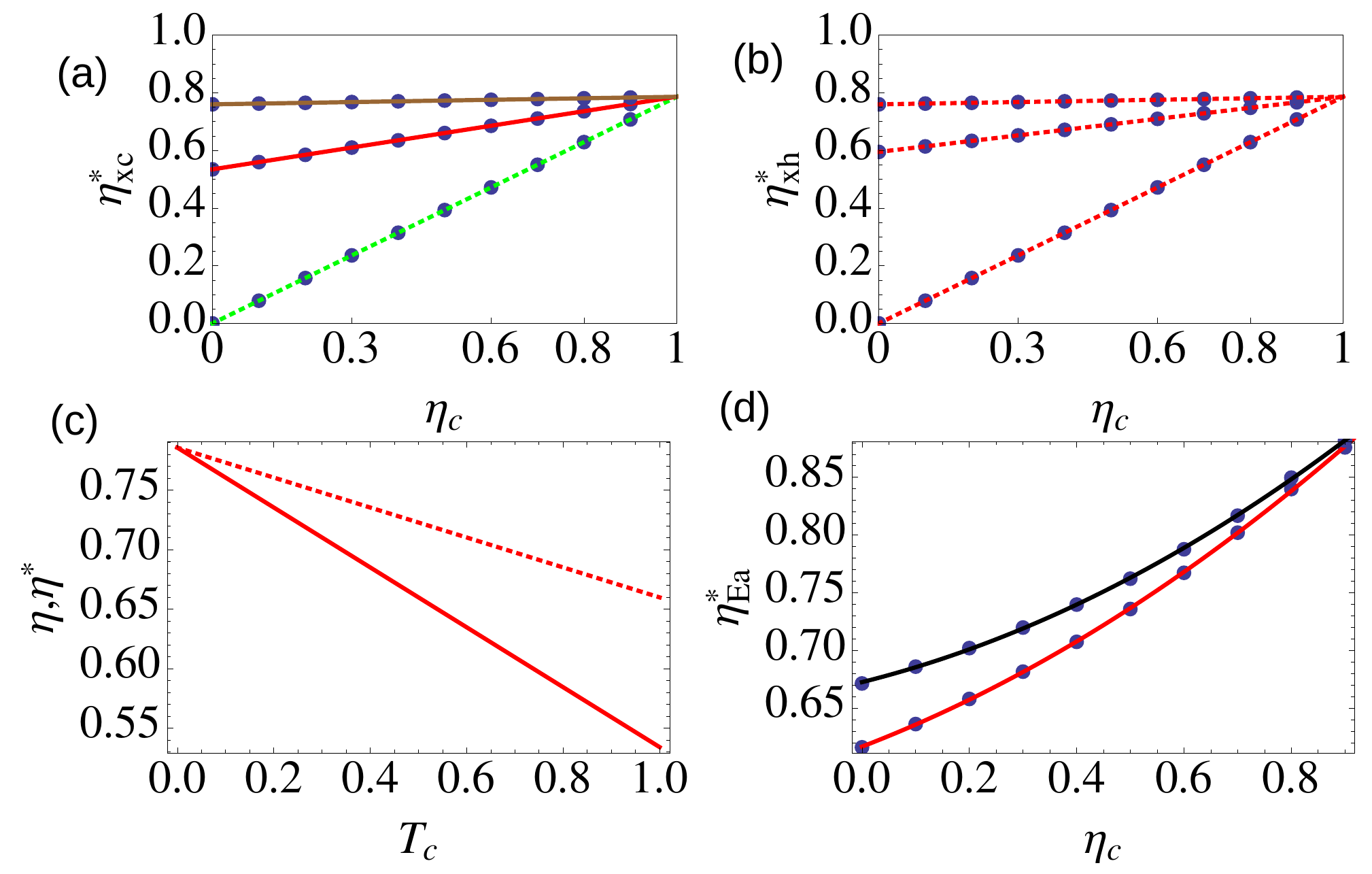}
\caption{ Linear dependence of $\eta_{xc}^*(a)$ and $\eta_{xh}^*(b)$ as a function of $\eta_C$, governed by Eq.(\ref{eq-empch}) evaluated at $x=\infty,1$ and $0$ (top to bottom ). Note that $\eta_{xh}^*=\eta_{xc}^*$ with the upper (middle) curves having a slope of $m=0.02(0.19)$ and intercept of $c=0.76 (0.59)$. c) Solid line represents the EMP, given by Eq.(\ref{eq-etaeax}) while the dotted line is simply the normal efficiency, $\eta =W/Q_h$. d) Appearance of a quadratic term and an intercept for $\eta_{E_a}^*$ as a function of $\eta_C$, evaluated at $x = 1.5(\infty)$ denoted by lower (upper) curves. The fit parameters for the upper (lower) curves are $a_1=0.85(0.76),a_2=1.7(0.75),a_3=0.05(-0.28),a_4=0.07(-0.28)$.}
\label{fig-emp-lin}
\end{figure}

In general, the EMP  has a  universally accepted formula, the Curzon-Ahlborn EMP, $\eta_{CA}=1-\sqrt{1-\eta_C}$ \cite{curzon1975efficiency,PhysRevLett.102.130602} and is represented by the dashed curves in Fig.(\ref{fig-emp}b,c and d). As a function of $\eta_C$, the EMP is bound between $\eta_C/2\le \eta^*\le \eta^{**}$, where the upper bound is $\eta^{**} =\frac{\eta_C}{2-\eta_C}$\cite{PhysRevLett.105.150603}.
In Fig.(\ref{fig-emp}b,c and d), we show the behavior of our engine's EMP as a function of the Carnot efficiency, $\eta_C$.  The solid (topmost green) curve represent the upper bound $\eta^{**}$. The EMP of the QHE optimized with respect to $x$ for $x_c=x_h=0$ is represented by the solid line highlighted with red dots. In Fig.(\ref{fig-emp}b,c), $\eta^*_x\ge(<)\eta_{CA}$ is observed under the condition $r\ge (<) g$. Values of EMP larger than $\eta_{CA}$ has been previously reported with squeezed reservoirs \cite{rossnagel2014nanoscale,klaers2017squeezed}. In our case, one can have EMP more than the predicted $\eta_{CA}$ just by squeezing the cavity even in the absence of squeezed reservoirs. In Fig.(\ref{fig-emp}d), for nonzero values of cavity-squeezing, $\eta_{E_a}^*> \eta_{CA}$ is shown (solid black curve). This result is valid irrespective of $r$ and $g$ values.   
The upper bound is always obeyed in presence of squeezing as evident from Fig.(\ref{fig-emp-lin}b,c and d). The EMP of the QHE is always lower than the upper dashed curve ($\eta^{**}$).  Note that the universal slope of $1/2$ (any $EMP=\eta_C/2$ near equilibrium)\cite{PhysRevLett.95.190602} is maintained in all the curves for smaller values of $\eta_C$ when maximized with respect to $x$. 

We now move to discuss a rather interesting finding observed when the EMP is maximized with respect to a reservoir squeezing parameter. As can be seen from Fig.(\ref{fig-emp-lin}a and b), both $\eta^*_{xh}$ and $\eta_{xc}^*$ are found to be linear in $\eta_C$ with a slope which is not equal to the universally predicted value of 1/2\cite{PhysRevLett.102.130602}. By a linear curve fitting technique, we infer that the EMP with respect to $x_c$ or $x_h$ is dictated by the equation,
\begin{align}
 \label{eq-empch}
 \eta_{xh}^* &= \eta_{xc}^*= 
 m \eta_C+c.
\end{align}
Our numerical results reveal that  the slope, $m$ is equal to the numerical value of $W_{diss}/Q_h$ and the intercept, $c$ being given by the numerical value of the quantity, $(E_{ab}-W_{diss})/Q_h$. This intercept is interestingly the efficiency of the engine albeit with $T_c=1$. Note that, $\eta_{xc}^*=\eta_{xh}^*$ and is shown as two identical plots in Fig.(\ref{fig-emp-lin}a,b). In these two figures. The numerical plots reveal that the $m\ne1/2$. Such a breakdown of the universality of the linear coefficient has also been observed in presence of geometric phaselike effects \cite{PhysRevE.99.022104,PhysRevE.106.024131}. Since $W_{diss}>1$, the EMP increases as $x$ is increased (for fixed $T_\ell$) to a maximum value of $E_{ab}/Q_h$ at $\eta_C=1$. The efficiency of the QHE, $\eta = W/Q_h$ is always less than $ \eta^*_\nu$ and is shown as a function of $T_c$ in Fig.(\ref{fig-emp-lin}c).

This linear dependence doesn't exist for $\eta_{E_a}^*$ for finite $x$ as seen from the numerical results in Fig.(\ref{fig-emp-lin}d) for $x=1$ and $x\to \infty$. It has been previously reported that such a nonlinear dependence of the EMP on the squeezing parameter $x$ takes the form $\eta^*_*=1-\sqrt{sech(2x)}\sqrt{1-\eta_C}$ \cite{liu2022optimized}. We assess the validity if this expression by defining two curve fitting equations,
\begin{align}
 \label{eq-etaeax}
 \eta_{E_a}^*&\approx  a_1-\sqrt{sech(a_2x)}\sqrt{a_3-a_4\eta_C}\\
 &\approx a_5\eta_C^{}+a_6\eta_C^2+c
 \label{eq-pol-fit}
\end{align}
 that can best represent the EMP with respect to the system parameter $E_a$. Here, $a_i$-s are fit parameters. We observe that $a_1\ne a_3\ne a_4\ne 1$ and $a_3\ne 2$ resulting in $\eta_{E_a}^*\ne \eta^*_*$ and is shown in Fig.(\ref{fig-emp-lin}d). Further, in Eq.(\ref{eq-pol-fit}), $a_5\ne 1/2$ and $a_6 \ne 1/8$. In this engine, it is already known that the quadratic coefficient is not $1/8$ \cite{PhysRevA.88.013842}. Both the above equations are good fits (solid curves) on the numerically evaluated $\eta_{E_a}^*$ (dots) as function of $\eta_C$ as seen in Fig.(\ref{fig-emp-lin}d). It is interesting to note that the intercept of $\eta_{E_a}^*$ as a function of $\eta_C$ in Eq.(\ref{eq-pol-fit}) is the same numerical value of the engine's efficiency of the engine, $\eta = W/Q_h$ similar to what was observed in Eq.(\ref{eq-empch}). This lets us rationalize that Eq.(\ref{eq-pol-fit}) is a better representation of $\eta_{E_a}^*$ vs $\eta_C$ than Eq.(\ref{eq-etaeax}).   At $\eta_C=1$, $\eta_{E_a}^*$ again reaches a maximum value of $E_{ab}/Q_h$.  For $x=0$, $m = 1/2$ is recovered. Further for $x=0$, the intercept in Eq.(\ref{eq-etaeax}) also vanishes by mixing with the quadratic term. Since we cannot derive analytical expressions for these coefficients, we demonstrated it this numerically shown as the bottom-most dotted line in Fig.(\ref{fig-emp}d)).

The EMP also has other interesting logarithmic expressions\cite{PhysRevE.98.052137,dechant2017underdamped,iyyappan2020efficiency}, one particularly claimed to be valid for squeezed states\cite{PhysRevE.100.052126}, $\eta_L^*=\eta_m^2/\{1-(1-\eta_m)\ln(1-\eta_m)\}$. $\eta_m
$ is a modified Carnot efficiency given by $\eta_m=1-T_c/T_{h}^m$. $T_h^m$ is a modified but fictitious reservoir temperature and is directly proportional to the energy  of the squeezed mode and inversely proportional to the logarithmic ratio of the squeezed mode's occupation factor. By an analogy with this previous work \cite{PhysRevE.100.052126}, we can express the modified temperature  in our QHE to be,
\begin{align}
 \label{eq-tm}
 T_h^m&=\displaystyle\frac{E_a-E_1}{\ln\frac{1+N_h}{N_h}}.
\end{align}

\begin{figure}
\centering
\includegraphics[width=8cm]{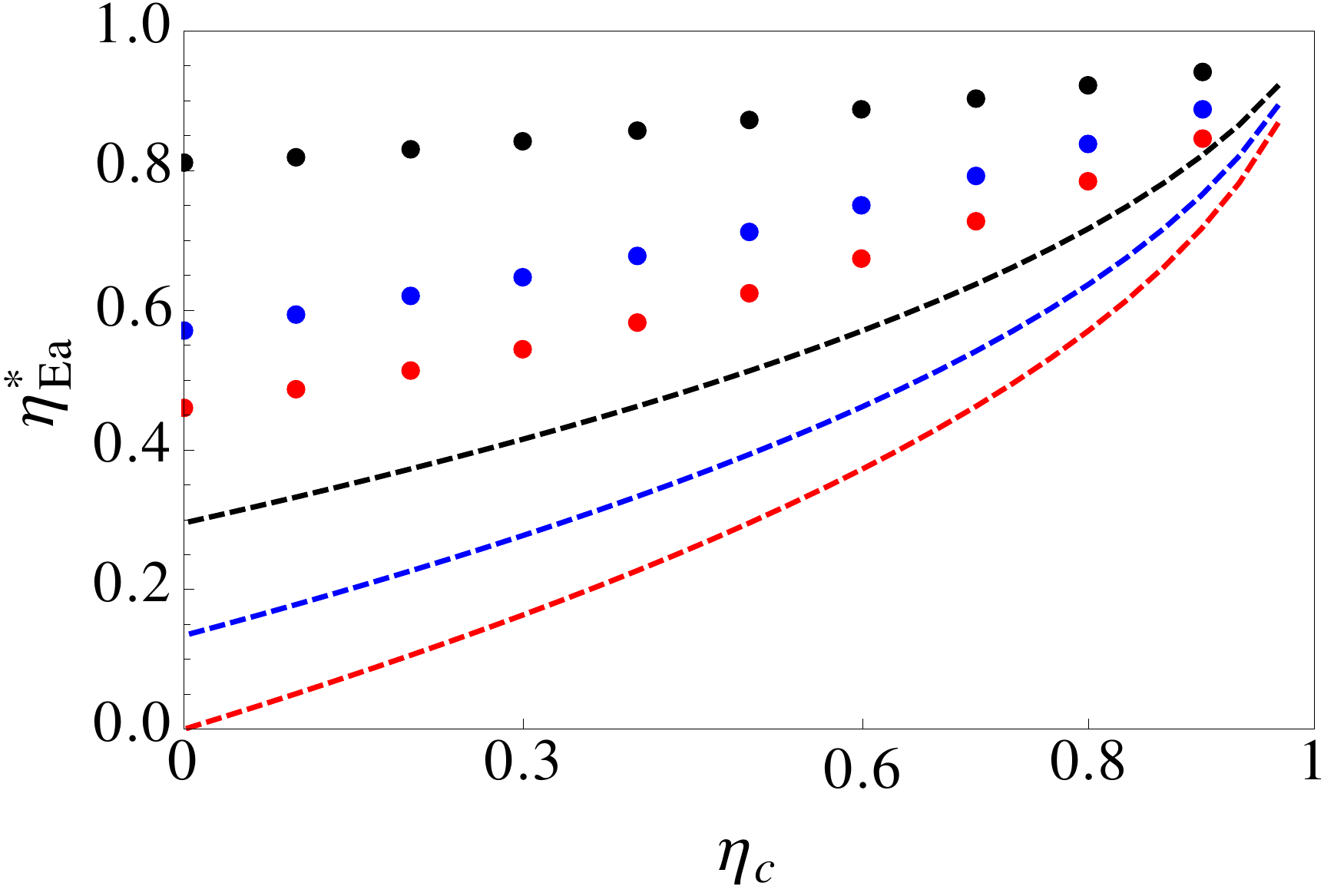}
\caption{Disagreement between the QHE's EMP optimized with respect to $E_a$ and the predicted EMP, $\eta_L^*$ for the same parameters. $\eta_L^*$ is evaluated using the definition in Eq.(\ref{eq-tm}). The dotted (dashed) curves represent $\eta_{E_a}^* (\eta_L^*)$. Parameters used are $T_h=3, x_c=x_h=0.1, x= 0.6$ (top dotted), $T_h=4, x_c=x_h=0.2, x= 0.5 $ (middle dotted) and $T_h=6, x_c=x_h=0.2, x= 2\pi$. }
\label{fig-emp-invalid}
\end{figure}

 We numerically evaluate $\eta_{E_a^*}$ for different squeezing parameters and $T_h$ values and plot it in Fig.(\ref{fig-emp-invalid}) along side the corresponding $\eta_L^*$ values. As can be seen, $\eta^*_{E_a}\ne \eta_L^*$. Further since $\eta_{xh}^*$ and $\eta_{xc}^*$ is found to be linear in $\eta_C$, these anyway don't agree with the predicted value $\eta_L^*$. Under extremely low squeezing conditions of the hot bath, $\eta_C^m\to \eta_C$ in the expression for $\eta_L^*$. Under this condition, $\eta_L^*$ has been high
lighted as dotted curves in Fig.(\ref{fig-emp}b,c and d) and is seen to be unequal to $\eta_x^*$.

\section{Conclusion}
By deriving a coherence-population coupled quantum master equation, we carried out a comprehensive study of the thermodynamics of quantum heat engine coupled to two squeezed reservoirs and a squeezed unimodal cavity. We showed that the steadystate value of the coherence term of the density matrix vanishes (saturates) under maximal squeezing of the cold (hot) bath. Under high squeezing conditions of the cavity, the two upper states of the engine equipopulate.  We showed that under high squeezing of the cavity, the quantum coherence can no longer optimize  the flux beyond the classical values. We also showed how the flux can be linearized with respect to coherences under high squeezing conditions and equal Bose-Einstein distributions for the hot and cold baths. We also showed that larger EMP favors lower values of cavity temperatures and lower values of squeezing. The EMP can be increased beyond the Curzon-Ahlborn limit by squeezing the cavity alone even if the  baths are unsqueezed. We also show a linear dependence of the EMP with respect to the reservoirs' squeezing parameters which we identify analytically with a slope proportional to the dissipation into the cavity mode. The EMP with respect to a system parameter, $E_a$ doesn't obey the universal slope of $1/2$ for finite squeezing and is not equal to a recently proposed general form of the EMP in presence of squeezed reservoirs \cite{PhysRevE.100.052126}.

\begin{acknowledgments}
  MJS and HPG acknowledge the support from Science and Engineering Board, India for the start-up grant, SERB/SRG/2021/001088. 
\end{acknowledgments}

\bibliography{references.bib}

\begin{thebibliography}{58}%
\makeatletter
\providecommand \@ifxundefined [1]{%
 \@ifx{#1\undefined}
}%
\providecommand \@ifnum [1]{%
 \ifnum #1\expandafter \@firstoftwo
 \else \expandafter \@secondoftwo
 \fi
}%
\providecommand \@ifx [1]{%
 \ifx #1\expandafter \@firstoftwo
 \else \expandafter \@secondoftwo
 \fi
}%
\providecommand \natexlab [1]{#1}%
\providecommand \enquote  [1]{``#1''}%
\providecommand \bibnamefont  [1]{#1}%
\providecommand \bibfnamefont [1]{#1}%
\providecommand \citenamefont [1]{#1}%
\providecommand \href@noop [0]{\@secondoftwo}%
\providecommand \href [0]{\begingroup \@sanitize@url \@href}%
\providecommand \@href[1]{\@@startlink{#1}\@@href}%
\providecommand \@@href[1]{\endgroup#1\@@endlink}%
\providecommand \@sanitize@url [0]{\catcode `\\12\catcode `\$12\catcode
  `\&12\catcode `\#12\catcode `\^12\catcode `\_12\catcode `\%12\relax}%
\providecommand \@@startlink[1]{}%
\providecommand \@@endlink[0]{}%
\providecommand \url  [0]{\begingroup\@sanitize@url \@url }%
\providecommand \@url [1]{\endgroup\@href {#1}{\urlprefix }}%
\providecommand \urlprefix  [0]{URL }%
\providecommand \Eprint [0]{\href }%
\providecommand \doibase [0]{http://dx.doi.org/}%
\providecommand \selectlanguage [0]{\@gobble}%
\providecommand \bibinfo  [0]{\@secondoftwo}%
\providecommand \bibfield  [0]{\@secondoftwo}%
\providecommand \translation [1]{[#1]}%
\providecommand \BibitemOpen [0]{}%
\providecommand \bibitemStop [0]{}%
\providecommand \bibitemNoStop [0]{.\EOS\space}%
\providecommand \EOS [0]{\spacefactor3000\relax}%
\providecommand \BibitemShut  [1]{\csname bibitem#1\endcsname}%
\let\auto@bib@innerbib\@empty
\bibitem [{\citenamefont {Quan}\ \emph {et~al.}(2007)\citenamefont {Quan},
  \citenamefont {Liu}, \citenamefont {Sun},\ and\ \citenamefont
  {Nori}}]{quan2007quantum}%
  \BibitemOpen
  \bibfield  {author} {\bibinfo {author} {\bibfnamefont {H.-T.}\ \bibnamefont
  {Quan}}, \bibinfo {author} {\bibfnamefont {Y.-x.}\ \bibnamefont {Liu}},
  \bibinfo {author} {\bibfnamefont {C.-P.}\ \bibnamefont {Sun}}, \ and\
  \bibinfo {author} {\bibfnamefont {F.}~\bibnamefont {Nori}},\ }\href@noop {}
  {\bibfield  {journal} {\bibinfo  {journal} {Physical Review E}\ }\textbf
  {\bibinfo {volume} {76}},\ \bibinfo {pages} {031105} (\bibinfo {year}
  {2007})}\BibitemShut {NoStop}%
\bibitem [{\citenamefont {Kosloff}\ and\ \citenamefont
  {Levy}(2014)}]{kosloff2014quantum}%
  \BibitemOpen
  \bibfield  {author} {\bibinfo {author} {\bibfnamefont {R.}~\bibnamefont
  {Kosloff}}\ and\ \bibinfo {author} {\bibfnamefont {A.}~\bibnamefont {Levy}},\
  }\href@noop {} {\bibfield  {journal} {\bibinfo  {journal} {Annual Review of
  Physical Chemistry}\ }\textbf {\bibinfo {volume} {65}},\ \bibinfo {pages}
  {365} (\bibinfo {year} {2014})}\BibitemShut {NoStop}%
\bibitem [{\citenamefont {Campisi}\ \emph {et~al.}(2015)\citenamefont
  {Campisi}, \citenamefont {Pekola},\ and\ \citenamefont
  {Fazio}}]{campisi2015nonequilibrium}%
  \BibitemOpen
  \bibfield  {author} {\bibinfo {author} {\bibfnamefont {M.}~\bibnamefont
  {Campisi}}, \bibinfo {author} {\bibfnamefont {J.}~\bibnamefont {Pekola}}, \
  and\ \bibinfo {author} {\bibfnamefont {R.}~\bibnamefont {Fazio}},\
  }\href@noop {} {\bibfield  {journal} {\bibinfo  {journal} {New Journal of
  Physics}\ }\textbf {\bibinfo {volume} {17}},\ \bibinfo {pages} {035012}
  (\bibinfo {year} {2015})}\BibitemShut {NoStop}%
\bibitem [{\citenamefont {Scovil}\ and\ \citenamefont
  {Schulz-DuBois}(1959)}]{PhysRevLett.2.262}%
  \BibitemOpen
  \bibfield  {author} {\bibinfo {author} {\bibfnamefont {H.}~\bibnamefont
  {Scovil}}\ and\ \bibinfo {author} {\bibfnamefont {E.}~\bibnamefont
  {Schulz-DuBois}},\ }\href@noop {} {\bibfield  {journal} {\bibinfo  {journal}
  {Phys. Rev. Lett.}\ }\textbf {\bibinfo {volume} {2}},\ \bibinfo {pages} {262}
  (\bibinfo {year} {1959})}\BibitemShut {NoStop}%
\bibitem [{\citenamefont {Brantut}\ \emph {et~al.}(2013)\citenamefont
  {Brantut}, \citenamefont {Grenier}, \citenamefont {Meineke}, \citenamefont
  {Stadler}, \citenamefont {Krinner}, \citenamefont {Kollath}, \citenamefont
  {Esslinger},\ and\ \citenamefont {Georges}}]{science.342.713}%
  \BibitemOpen
  \bibfield  {author} {\bibinfo {author} {\bibfnamefont {J.-P.}\ \bibnamefont
  {Brantut}}, \bibinfo {author} {\bibfnamefont {C.}~\bibnamefont {Grenier}},
  \bibinfo {author} {\bibfnamefont {J.}~\bibnamefont {Meineke}}, \bibinfo
  {author} {\bibfnamefont {D.}~\bibnamefont {Stadler}}, \bibinfo {author}
  {\bibfnamefont {S.}~\bibnamefont {Krinner}}, \bibinfo {author} {\bibfnamefont
  {C.}~\bibnamefont {Kollath}}, \bibinfo {author} {\bibfnamefont
  {T.}~\bibnamefont {Esslinger}}, \ and\ \bibinfo {author} {\bibfnamefont
  {A.}~\bibnamefont {Georges}},\ }\href@noop {} {\bibfield  {journal} {\bibinfo
   {journal} {Science}\ }\textbf {\bibinfo {volume} {342}},\ \bibinfo {pages}
  {713} (\bibinfo {year} {2013})}\BibitemShut {NoStop}%
\bibitem [{\citenamefont {Klatzow}\ \emph {et~al.}(2019)\citenamefont
  {Klatzow}, \citenamefont {Becker}, \citenamefont {Ledingham}, \citenamefont
  {Weinzetl}, \citenamefont {Kaczmarek}, \citenamefont {Saunders},
  \citenamefont {Nunn}, \citenamefont {Walmsley}, \citenamefont {Uzdin},\ and\
  \citenamefont {Poem}}]{PhysRevLett.122.110601}%
  \BibitemOpen
  \bibfield  {author} {\bibinfo {author} {\bibfnamefont {J.}~\bibnamefont
  {Klatzow}}, \bibinfo {author} {\bibfnamefont {J.~N.}\ \bibnamefont {Becker}},
  \bibinfo {author} {\bibfnamefont {P.~M.}\ \bibnamefont {Ledingham}}, \bibinfo
  {author} {\bibfnamefont {C.}~\bibnamefont {Weinzetl}}, \bibinfo {author}
  {\bibfnamefont {K.~T.}\ \bibnamefont {Kaczmarek}}, \bibinfo {author}
  {\bibfnamefont {D.~J.}\ \bibnamefont {Saunders}}, \bibinfo {author}
  {\bibfnamefont {J.}~\bibnamefont {Nunn}}, \bibinfo {author} {\bibfnamefont
  {I.~A.}\ \bibnamefont {Walmsley}}, \bibinfo {author} {\bibfnamefont
  {R.}~\bibnamefont {Uzdin}}, \ and\ \bibinfo {author} {\bibfnamefont
  {E.}~\bibnamefont {Poem}},\ }\href@noop {} {\bibfield  {journal} {\bibinfo
  {journal} {Phys. Rev. Lett.}\ }\textbf {\bibinfo {volume} {122}},\ \bibinfo
  {pages} {110601} (\bibinfo {year} {2019})}\BibitemShut {NoStop}%
\bibitem [{\citenamefont {Scully}\ \emph {et~al.}(2011)\citenamefont {Scully},
  \citenamefont {Chapin}, \citenamefont {Dorfman}, \citenamefont {Kim},\ and\
  \citenamefont {Svidzinsky}}]{PNAS.108.15097}%
  \BibitemOpen
  \bibfield  {author} {\bibinfo {author} {\bibfnamefont {M.~O.}\ \bibnamefont
  {Scully}}, \bibinfo {author} {\bibfnamefont {K.~R.}\ \bibnamefont {Chapin}},
  \bibinfo {author} {\bibfnamefont {K.~E.}\ \bibnamefont {Dorfman}}, \bibinfo
  {author} {\bibfnamefont {M.~B.}\ \bibnamefont {Kim}}, \ and\ \bibinfo
  {author} {\bibfnamefont {A.}~\bibnamefont {Svidzinsky}},\ }\href {\doibase
  10.1073/pnas.1110234108} {\bibfield  {journal} {\bibinfo  {journal} {Proc.
  Natl. Acad. Sci. U.S.A.}\ }\textbf {\bibinfo {volume} {108}},\ \bibinfo
  {pages} {15097} (\bibinfo {year} {2011})}\BibitemShut {NoStop}%
\bibitem [{\citenamefont {Scully}\ \emph {et~al.}(2003)\citenamefont {Scully},
  \citenamefont {Zubairy}, \citenamefont {Agarwal},\ and\ \citenamefont
  {Walther}}]{doi:10.1126/science.1078955}%
  \BibitemOpen
  \bibfield  {author} {\bibinfo {author} {\bibfnamefont {M.~O.}\ \bibnamefont
  {Scully}}, \bibinfo {author} {\bibfnamefont {M.~S.}\ \bibnamefont {Zubairy}},
  \bibinfo {author} {\bibfnamefont {G.~S.}\ \bibnamefont {Agarwal}}, \ and\
  \bibinfo {author} {\bibfnamefont {H.}~\bibnamefont {Walther}},\ }\href
  {\doibase 10.1126/science.1078955} {\bibfield  {journal} {\bibinfo  {journal}
  {Science}\ }\textbf {\bibinfo {volume} {299}},\ \bibinfo {pages} {862}
  (\bibinfo {year} {2003})},\ \Eprint
  {http://arxiv.org/abs/https://www.science.org/doi/pdf/10.1126/science.1078955}
  {https://www.science.org/doi/pdf/10.1126/science.1078955} \BibitemShut
  {NoStop}%
\bibitem [{\citenamefont {Huang}\ \emph {et~al.}(2012)\citenamefont {Huang},
  \citenamefont {Wang},\ and\ \citenamefont {Yi}}]{PhysRevE.86.051105}%
  \BibitemOpen
  \bibfield  {author} {\bibinfo {author} {\bibfnamefont {X.~L.}\ \bibnamefont
  {Huang}}, \bibinfo {author} {\bibfnamefont {T.}~\bibnamefont {Wang}}, \ and\
  \bibinfo {author} {\bibfnamefont {X.~X.}\ \bibnamefont {Yi}},\ }\href
  {\doibase 10.1103/PhysRevE.86.051105} {\bibfield  {journal} {\bibinfo
  {journal} {Phys. Rev. E}\ }\textbf {\bibinfo {volume} {86}},\ \bibinfo
  {pages} {051105} (\bibinfo {year} {2012})}\BibitemShut {NoStop}%
\bibitem [{\citenamefont {Manzano}\ \emph
  {et~al.}(2016{\natexlab{a}})\citenamefont {Manzano}, \citenamefont {Galve},
  \citenamefont {Zambrini},\ and\ \citenamefont
  {Parrondo}}]{PhysRevE.93.052120}%
  \BibitemOpen
  \bibfield  {author} {\bibinfo {author} {\bibfnamefont {G.}~\bibnamefont
  {Manzano}}, \bibinfo {author} {\bibfnamefont {F.}~\bibnamefont {Galve}},
  \bibinfo {author} {\bibfnamefont {R.}~\bibnamefont {Zambrini}}, \ and\
  \bibinfo {author} {\bibfnamefont {J.~M.~R.}\ \bibnamefont {Parrondo}},\
  }\href {\doibase 10.1103/PhysRevE.93.052120} {\bibfield  {journal} {\bibinfo
  {journal} {Phys. Rev. E}\ }\textbf {\bibinfo {volume} {93}},\ \bibinfo
  {pages} {052120} (\bibinfo {year} {2016}{\natexlab{a}})}\BibitemShut
  {NoStop}%
\bibitem [{\citenamefont {Wang}\ \emph {et~al.}(2019)\citenamefont {Wang},
  \citenamefont {He},\ and\ \citenamefont {Ma}}]{PhysRevE.100.052126}%
  \BibitemOpen
  \bibfield  {author} {\bibinfo {author} {\bibfnamefont {J.}~\bibnamefont
  {Wang}}, \bibinfo {author} {\bibfnamefont {J.}~\bibnamefont {He}}, \ and\
  \bibinfo {author} {\bibfnamefont {Y.}~\bibnamefont {Ma}},\ }\href {\doibase
  10.1103/PhysRevE.100.052126} {\bibfield  {journal} {\bibinfo  {journal}
  {Phys. Rev. E}\ }\textbf {\bibinfo {volume} {100}},\ \bibinfo {pages}
  {052126} (\bibinfo {year} {2019})}\BibitemShut {NoStop}%
\bibitem [{\citenamefont {Manzano}(2018)}]{PhysRevE.98.042123}%
  \BibitemOpen
  \bibfield  {author} {\bibinfo {author} {\bibfnamefont {G.}~\bibnamefont
  {Manzano}},\ }\href {\doibase 10.1103/PhysRevE.98.042123} {\bibfield
  {journal} {\bibinfo  {journal} {Phys. Rev. E}\ }\textbf {\bibinfo {volume}
  {98}},\ \bibinfo {pages} {042123} (\bibinfo {year} {2018})}\BibitemShut
  {NoStop}%
\bibitem [{\citenamefont {Walls}(1983)}]{walls1983squeezed}%
  \BibitemOpen
  \bibfield  {author} {\bibinfo {author} {\bibfnamefont {D.~F.}\ \bibnamefont
  {Walls}},\ }\href@noop {} {\bibfield  {journal} {\bibinfo  {journal}
  {nature}\ }\textbf {\bibinfo {volume} {306}},\ \bibinfo {pages} {141}
  (\bibinfo {year} {1983})}\BibitemShut {NoStop}%
\bibitem [{\citenamefont {Puri}(1997)}]{puri1997coherent}%
  \BibitemOpen
  \bibfield  {author} {\bibinfo {author} {\bibfnamefont {R.}~\bibnamefont
  {Puri}},\ }\href@noop {} {\bibfield  {journal} {\bibinfo  {journal}
  {pramana}\ }\textbf {\bibinfo {volume} {48}},\ \bibinfo {pages} {787}
  (\bibinfo {year} {1997})}\BibitemShut {NoStop}%
\bibitem [{\citenamefont {Dupays}\ and\ \citenamefont
  {Chenu}(2021)}]{dupays2021shortcuts}%
  \BibitemOpen
  \bibfield  {author} {\bibinfo {author} {\bibfnamefont {L.}~\bibnamefont
  {Dupays}}\ and\ \bibinfo {author} {\bibfnamefont {A.}~\bibnamefont {Chenu}},\
  }\href@noop {} {\bibfield  {journal} {\bibinfo  {journal} {Quantum}\ }\textbf
  {\bibinfo {volume} {5}},\ \bibinfo {pages} {449} (\bibinfo {year}
  {2021})}\BibitemShut {NoStop}%
\bibitem [{\citenamefont {Kumar}\ \emph {et~al.}(2022)\citenamefont {Kumar},
  \citenamefont {Bagarti}, \citenamefont {Lahiri},\ and\ \citenamefont
  {Banerjee}}]{kumar2022thermodynamics}%
  \BibitemOpen
  \bibfield  {author} {\bibinfo {author} {\bibfnamefont {A.}~\bibnamefont
  {Kumar}}, \bibinfo {author} {\bibfnamefont {T.}~\bibnamefont {Bagarti}},
  \bibinfo {author} {\bibfnamefont {S.}~\bibnamefont {Lahiri}}, \ and\ \bibinfo
  {author} {\bibfnamefont {S.}~\bibnamefont {Banerjee}},\ }\href@noop {}
  {\bibfield  {journal} {\bibinfo  {journal} {arXiv preprint arXiv:2209.06433}\
  } (\bibinfo {year} {2022})}\BibitemShut {NoStop}%
\bibitem [{\citenamefont {Klaers}\ \emph
  {et~al.}(2017{\natexlab{a}})\citenamefont {Klaers}, \citenamefont {Faelt},
  \citenamefont {Imamoglu},\ and\ \citenamefont {Togan}}]{klaers2017squeezed}%
  \BibitemOpen
  \bibfield  {author} {\bibinfo {author} {\bibfnamefont {J.}~\bibnamefont
  {Klaers}}, \bibinfo {author} {\bibfnamefont {S.}~\bibnamefont {Faelt}},
  \bibinfo {author} {\bibfnamefont {A.}~\bibnamefont {Imamoglu}}, \ and\
  \bibinfo {author} {\bibfnamefont {E.}~\bibnamefont {Togan}},\ }\href@noop {}
  {\bibfield  {journal} {\bibinfo  {journal} {Physical Review X}\ }\textbf
  {\bibinfo {volume} {7}},\ \bibinfo {pages} {031044} (\bibinfo {year}
  {2017}{\natexlab{a}})}\BibitemShut {NoStop}%
\bibitem [{\citenamefont {Klaers}\ \emph
  {et~al.}(2017{\natexlab{b}})\citenamefont {Klaers}, \citenamefont {Faelt},
  \citenamefont {Imamoglu},\ and\ \citenamefont {Togan}}]{PhysRevX.7.031044}%
  \BibitemOpen
  \bibfield  {author} {\bibinfo {author} {\bibfnamefont {J.}~\bibnamefont
  {Klaers}}, \bibinfo {author} {\bibfnamefont {S.}~\bibnamefont {Faelt}},
  \bibinfo {author} {\bibfnamefont {A.}~\bibnamefont {Imamoglu}}, \ and\
  \bibinfo {author} {\bibfnamefont {E.}~\bibnamefont {Togan}},\ }\href
  {\doibase 10.1103/PhysRevX.7.031044} {\bibfield  {journal} {\bibinfo
  {journal} {Phys. Rev. X}\ }\textbf {\bibinfo {volume} {7}},\ \bibinfo {pages}
  {031044} (\bibinfo {year} {2017}{\natexlab{b}})}\BibitemShut {NoStop}%
\bibitem [{\citenamefont {Pal}\ \emph {et~al.}(2019)\citenamefont {Pal},
  \citenamefont {Mahesh},\ and\ \citenamefont
  {Agarwalla}}]{PhysRevA.100.042119}%
  \BibitemOpen
  \bibfield  {author} {\bibinfo {author} {\bibfnamefont {S.}~\bibnamefont
  {Pal}}, \bibinfo {author} {\bibfnamefont {T.}~\bibnamefont {Mahesh}}, \ and\
  \bibinfo {author} {\bibfnamefont {B.~K.}\ \bibnamefont {Agarwalla}},\
  }\href@noop {} {\bibfield  {journal} {\bibinfo  {journal} {Physical Review
  A}\ }\textbf {\bibinfo {volume} {100}},\ \bibinfo {pages} {042119} (\bibinfo
  {year} {2019})}\BibitemShut {NoStop}%
\bibitem [{\citenamefont {Ro{\ss}nagel}\ \emph {et~al.}(2014)\citenamefont
  {Ro{\ss}nagel}, \citenamefont {Abah}, \citenamefont {Schmidt-Kaler},
  \citenamefont {Singer},\ and\ \citenamefont {Lutz}}]{rossnagel2014nanoscale}%
  \BibitemOpen
  \bibfield  {author} {\bibinfo {author} {\bibfnamefont {J.}~\bibnamefont
  {Ro{\ss}nagel}}, \bibinfo {author} {\bibfnamefont {O.}~\bibnamefont {Abah}},
  \bibinfo {author} {\bibfnamefont {F.}~\bibnamefont {Schmidt-Kaler}}, \bibinfo
  {author} {\bibfnamefont {K.}~\bibnamefont {Singer}}, \ and\ \bibinfo {author}
  {\bibfnamefont {E.}~\bibnamefont {Lutz}},\ }\href@noop {} {\bibfield
  {journal} {\bibinfo  {journal} {Physical review letters}\ }\textbf {\bibinfo
  {volume} {112}},\ \bibinfo {pages} {030602} (\bibinfo {year}
  {2014})}\BibitemShut {NoStop}%
\bibitem [{\citenamefont {Zou}\ \emph {et~al.}(2017)\citenamefont {Zou},
  \citenamefont {Jiang}, \citenamefont {Mei}, \citenamefont {Guo},\ and\
  \citenamefont {Du}}]{zou2017quantum}%
  \BibitemOpen
  \bibfield  {author} {\bibinfo {author} {\bibfnamefont {Y.}~\bibnamefont
  {Zou}}, \bibinfo {author} {\bibfnamefont {Y.}~\bibnamefont {Jiang}}, \bibinfo
  {author} {\bibfnamefont {Y.}~\bibnamefont {Mei}}, \bibinfo {author}
  {\bibfnamefont {X.}~\bibnamefont {Guo}}, \ and\ \bibinfo {author}
  {\bibfnamefont {S.}~\bibnamefont {Du}},\ }\href@noop {} {\bibfield  {journal}
  {\bibinfo  {journal} {Physical Review Letters}\ }\textbf {\bibinfo {volume}
  {119}},\ \bibinfo {pages} {050602} (\bibinfo {year} {2017})}\BibitemShut
  {NoStop}%
\bibitem [{\citenamefont {Melo}\ \emph {et~al.}(2022)\citenamefont {Melo},
  \citenamefont {S{\'a}}, \citenamefont {Roditi}, \citenamefont {Sarthour},
  \citenamefont {Oliveira},\ and\ \citenamefont
  {Souza}}]{melo2022experimental}%
  \BibitemOpen
  \bibfield  {author} {\bibinfo {author} {\bibfnamefont {F.~V.}\ \bibnamefont
  {Melo}}, \bibinfo {author} {\bibfnamefont {N.}~\bibnamefont {S{\'a}}},
  \bibinfo {author} {\bibfnamefont {I.}~\bibnamefont {Roditi}}, \bibinfo
  {author} {\bibfnamefont {R.~S.}\ \bibnamefont {Sarthour}}, \bibinfo {author}
  {\bibfnamefont {I.~S.}\ \bibnamefont {Oliveira}}, \ and\ \bibinfo {author}
  {\bibfnamefont {A.~M.}\ \bibnamefont {Souza}},\ }\href@noop {} {\bibfield
  {journal} {\bibinfo  {journal} {arXiv preprint arXiv:2203.13773}\ } (\bibinfo
  {year} {2022})}\BibitemShut {NoStop}%
\bibitem [{\citenamefont {Niedenzu}\ \emph {et~al.}(2016)\citenamefont
  {Niedenzu}, \citenamefont {Gelbwaser-Klimovsky}, \citenamefont {Kofman},\
  and\ \citenamefont {Kurizki}}]{Niedenzu_2016}%
  \BibitemOpen
  \bibfield  {author} {\bibinfo {author} {\bibfnamefont {W.}~\bibnamefont
  {Niedenzu}}, \bibinfo {author} {\bibfnamefont {D.}~\bibnamefont
  {Gelbwaser-Klimovsky}}, \bibinfo {author} {\bibfnamefont {A.~G.}\
  \bibnamefont {Kofman}}, \ and\ \bibinfo {author} {\bibfnamefont
  {G.}~\bibnamefont {Kurizki}},\ }\href {\doibase
  10.1088/1367-2630/18/8/083012} {\bibfield  {journal} {\bibinfo  {journal}
  {New Journal of Physics}\ }\textbf {\bibinfo {volume} {18}},\ \bibinfo
  {pages} {083012} (\bibinfo {year} {2016})}\BibitemShut {NoStop}%
\bibitem [{\citenamefont {Lostaglio}\ \emph
  {et~al.}(2015{\natexlab{a}})\citenamefont {Lostaglio}, \citenamefont
  {Jennings},\ and\ \citenamefont {Rudolph}}]{Lostaglio_2015}%
  \BibitemOpen
  \bibfield  {author} {\bibinfo {author} {\bibfnamefont {M.}~\bibnamefont
  {Lostaglio}}, \bibinfo {author} {\bibfnamefont {D.}~\bibnamefont {Jennings}},
  \ and\ \bibinfo {author} {\bibfnamefont {T.}~\bibnamefont {Rudolph}},\ }\href
  {\doibase 10.1038/ncomms7383} {\bibfield  {journal} {\bibinfo  {journal}
  {Nature Communications}\ }\textbf {\bibinfo {volume} {6}} (\bibinfo {year}
  {2015}{\natexlab{a}}),\ 10.1038/ncomms7383}\BibitemShut {NoStop}%
\bibitem [{\citenamefont {Lostaglio}\ \emph
  {et~al.}(2015{\natexlab{b}})\citenamefont {Lostaglio}, \citenamefont
  {Korzekwa}, \citenamefont {Jennings},\ and\ \citenamefont
  {Rudolph}}]{PhysRevX.5.021001}%
  \BibitemOpen
  \bibfield  {author} {\bibinfo {author} {\bibfnamefont {M.}~\bibnamefont
  {Lostaglio}}, \bibinfo {author} {\bibfnamefont {K.}~\bibnamefont {Korzekwa}},
  \bibinfo {author} {\bibfnamefont {D.}~\bibnamefont {Jennings}}, \ and\
  \bibinfo {author} {\bibfnamefont {T.}~\bibnamefont {Rudolph}},\ }\href
  {\doibase 10.1103/PhysRevX.5.021001} {\bibfield  {journal} {\bibinfo
  {journal} {Phys. Rev. X}\ }\textbf {\bibinfo {volume} {5}},\ \bibinfo {pages}
  {021001} (\bibinfo {year} {2015}{\natexlab{b}})}\BibitemShut {NoStop}%
\bibitem [{\citenamefont {Korzekwa}\ \emph {et~al.}(2016)\citenamefont
  {Korzekwa}, \citenamefont {Lostaglio}, \citenamefont {Oppenheim},\ and\
  \citenamefont {Jennings}}]{Korzekwa_2016}%
  \BibitemOpen
  \bibfield  {author} {\bibinfo {author} {\bibfnamefont {K.}~\bibnamefont
  {Korzekwa}}, \bibinfo {author} {\bibfnamefont {M.}~\bibnamefont {Lostaglio}},
  \bibinfo {author} {\bibfnamefont {J.}~\bibnamefont {Oppenheim}}, \ and\
  \bibinfo {author} {\bibfnamefont {D.}~\bibnamefont {Jennings}},\ }\href
  {\doibase 10.1088/1367-2630/18/2/023045} {\bibfield  {journal} {\bibinfo
  {journal} {New Journal of Physics}\ }\textbf {\bibinfo {volume} {18}},\
  \bibinfo {pages} {023045} (\bibinfo {year} {2016})}\BibitemShut {NoStop}%
\bibitem [{\citenamefont {Abah}\ and\ \citenamefont {Lutz}(2014)}]{Abah_2014}%
  \BibitemOpen
  \bibfield  {author} {\bibinfo {author} {\bibfnamefont {O.}~\bibnamefont
  {Abah}}\ and\ \bibinfo {author} {\bibfnamefont {E.}~\bibnamefont {Lutz}},\
  }\href {\doibase 10.1209/0295-5075/106/20001} {\bibfield  {journal} {\bibinfo
   {journal} {{EPL} (Europhysics Letters)}\ }\textbf {\bibinfo {volume}
  {106}},\ \bibinfo {pages} {20001} (\bibinfo {year} {2014})}\BibitemShut
  {NoStop}%
\bibitem [{\citenamefont {Ro\ss{}nagel}\ \emph {et~al.}(2014)\citenamefont
  {Ro\ss{}nagel}, \citenamefont {Abah}, \citenamefont {Schmidt-Kaler},
  \citenamefont {Singer},\ and\ \citenamefont {Lutz}}]{PhysRevLett.112.030602}%
  \BibitemOpen
  \bibfield  {author} {\bibinfo {author} {\bibfnamefont {J.}~\bibnamefont
  {Ro\ss{}nagel}}, \bibinfo {author} {\bibfnamefont {O.}~\bibnamefont {Abah}},
  \bibinfo {author} {\bibfnamefont {F.}~\bibnamefont {Schmidt-Kaler}}, \bibinfo
  {author} {\bibfnamefont {K.}~\bibnamefont {Singer}}, \ and\ \bibinfo {author}
  {\bibfnamefont {E.}~\bibnamefont {Lutz}},\ }\href {\doibase
  10.1103/PhysRevLett.112.030602} {\bibfield  {journal} {\bibinfo  {journal}
  {Phys. Rev. Lett.}\ }\textbf {\bibinfo {volume} {112}},\ \bibinfo {pages}
  {030602} (\bibinfo {year} {2014})}\BibitemShut {NoStop}%
\bibitem [{\citenamefont {Um}\ \emph {et~al.}(2022)\citenamefont {Um},
  \citenamefont {Dorfman},\ and\ \citenamefont {Park}}]{um2022coherence}%
  \BibitemOpen
  \bibfield  {author} {\bibinfo {author} {\bibfnamefont {J.}~\bibnamefont
  {Um}}, \bibinfo {author} {\bibfnamefont {K.~E.}\ \bibnamefont {Dorfman}}, \
  and\ \bibinfo {author} {\bibfnamefont {H.}~\bibnamefont {Park}},\ }\href@noop
  {} {\bibfield  {journal} {\bibinfo  {journal} {Physical Review Research}\
  }\textbf {\bibinfo {volume} {4}},\ \bibinfo {pages} {L032034} (\bibinfo
  {year} {2022})}\BibitemShut {NoStop}%
\bibitem [{\citenamefont {Goswami}\ and\ \citenamefont
  {Harbola}(2013)}]{PhysRevA.88.013842}%
  \BibitemOpen
  \bibfield  {author} {\bibinfo {author} {\bibfnamefont {H.~P.}\ \bibnamefont
  {Goswami}}\ and\ \bibinfo {author} {\bibfnamefont {U.}~\bibnamefont
  {Harbola}},\ }\href {\doibase 10.1103/PhysRevA.88.013842} {\bibfield
  {journal} {\bibinfo  {journal} {Phys. Rev. A}\ }\textbf {\bibinfo {volume}
  {88}},\ \bibinfo {pages} {013842} (\bibinfo {year} {2013})}\BibitemShut
  {NoStop}%
\bibitem [{\citenamefont {Rahav}\ \emph {et~al.}(2012)\citenamefont {Rahav},
  \citenamefont {Harbola},\ and\ \citenamefont {Mukamel}}]{PhysRevA.86.043843}%
  \BibitemOpen
  \bibfield  {author} {\bibinfo {author} {\bibfnamefont {S.}~\bibnamefont
  {Rahav}}, \bibinfo {author} {\bibfnamefont {U.}~\bibnamefont {Harbola}}, \
  and\ \bibinfo {author} {\bibfnamefont {S.}~\bibnamefont {Mukamel}},\ }\href
  {\doibase 10.1103/PhysRevA.86.043843} {\bibfield  {journal} {\bibinfo
  {journal} {Phys. Rev. A}\ }\textbf {\bibinfo {volume} {86}},\ \bibinfo
  {pages} {043843} (\bibinfo {year} {2012})}\BibitemShut {NoStop}%
\bibitem [{\citenamefont {Latune}\ \emph {et~al.}(2021)\citenamefont {Latune},
  \citenamefont {Sinayskiy},\ and\ \citenamefont
  {Petruccione}}]{latune2021roles}%
  \BibitemOpen
  \bibfield  {author} {\bibinfo {author} {\bibfnamefont {C.~L.}\ \bibnamefont
  {Latune}}, \bibinfo {author} {\bibfnamefont {I.}~\bibnamefont {Sinayskiy}}, \
  and\ \bibinfo {author} {\bibfnamefont {F.}~\bibnamefont {Petruccione}},\
  }\href@noop {} {\bibfield  {journal} {\bibinfo  {journal} {The European
  Physical Journal Special Topics}\ }\textbf {\bibinfo {volume} {230}},\
  \bibinfo {pages} {841} (\bibinfo {year} {2021})}\BibitemShut {NoStop}%
\bibitem [{\citenamefont {Manzano}\ \emph
  {et~al.}(2016{\natexlab{b}})\citenamefont {Manzano}, \citenamefont {Galve},
  \citenamefont {Zambrini},\ and\ \citenamefont {Parrondo}}]{Manzano_2016}%
  \BibitemOpen
  \bibfield  {author} {\bibinfo {author} {\bibfnamefont {G.}~\bibnamefont
  {Manzano}}, \bibinfo {author} {\bibfnamefont {F.}~\bibnamefont {Galve}},
  \bibinfo {author} {\bibfnamefont {R.}~\bibnamefont {Zambrini}}, \ and\
  \bibinfo {author} {\bibfnamefont {J.~M.~R.}\ \bibnamefont {Parrondo}},\
  }\href {\doibase 10.1103/physreve.93.052120} {\bibfield  {journal} {\bibinfo
  {journal} {Physical Review E}\ }\textbf {\bibinfo {volume} {93}} (\bibinfo
  {year} {2016}{\natexlab{b}}),\ 10.1103/physreve.93.052120}\BibitemShut
  {NoStop}%
\bibitem [{\citenamefont {Agarwalla}\ \emph
  {et~al.}(2017{\natexlab{a}})\citenamefont {Agarwalla}, \citenamefont
  {Jiang},\ and\ \citenamefont
  {Segal}}]{https://doi.org/10.48550/arxiv.1706.06206}%
  \BibitemOpen
  \bibfield  {author} {\bibinfo {author} {\bibfnamefont {B.~K.}\ \bibnamefont
  {Agarwalla}}, \bibinfo {author} {\bibfnamefont {J.-H.}\ \bibnamefont
  {Jiang}}, \ and\ \bibinfo {author} {\bibfnamefont {D.}~\bibnamefont
  {Segal}},\ }\href {\doibase 10.48550/ARXIV.1706.06206} {\  (\bibinfo {year}
  {2017}{\natexlab{a}}),\ 10.48550/ARXIV.1706.06206}\BibitemShut {NoStop}%
\bibitem [{\citenamefont {Long}\ and\ \citenamefont
  {Liu}(2015)}]{PhysRevE.91.062137}%
  \BibitemOpen
  \bibfield  {author} {\bibinfo {author} {\bibfnamefont {R.}~\bibnamefont
  {Long}}\ and\ \bibinfo {author} {\bibfnamefont {W.}~\bibnamefont {Liu}},\
  }\href {\doibase 10.1103/PhysRevE.91.062137} {\bibfield  {journal} {\bibinfo
  {journal} {Phys. Rev. E}\ }\textbf {\bibinfo {volume} {91}},\ \bibinfo
  {pages} {062137} (\bibinfo {year} {2015})}\BibitemShut {NoStop}%
\bibitem [{\citenamefont {Chen}\ \emph {et~al.}(2006)\citenamefont {Chen},
  \citenamefont {Church}, \citenamefont {Englert}, \citenamefont {Henkel},
  \citenamefont {Rohwedder}, \citenamefont {Scully},\ and\ \citenamefont
  {Zubairy}}]{chen2006quantum}%
  \BibitemOpen
  \bibfield  {author} {\bibinfo {author} {\bibfnamefont {G.}~\bibnamefont
  {Chen}}, \bibinfo {author} {\bibfnamefont {D.~A.}\ \bibnamefont {Church}},
  \bibinfo {author} {\bibfnamefont {B.-G.}\ \bibnamefont {Englert}}, \bibinfo
  {author} {\bibfnamefont {C.}~\bibnamefont {Henkel}}, \bibinfo {author}
  {\bibfnamefont {B.}~\bibnamefont {Rohwedder}}, \bibinfo {author}
  {\bibfnamefont {M.~O.}\ \bibnamefont {Scully}}, \ and\ \bibinfo {author}
  {\bibfnamefont {M.~S.}\ \bibnamefont {Zubairy}},\ }\href@noop {} {\emph
  {\bibinfo {title} {Quantum computing devices: principles, designs, and
  analysis}}}\ (\bibinfo  {publisher} {Chapman and Hall/CRC},\ \bibinfo {year}
  {2006})\BibitemShut {NoStop}%
\bibitem [{\citenamefont {Teich}\ and\ \citenamefont {Saleh}(1989)}]{article}%
  \BibitemOpen
  \bibfield  {author} {\bibinfo {author} {\bibfnamefont {M.}~\bibnamefont
  {Teich}}\ and\ \bibinfo {author} {\bibfnamefont {B.}~\bibnamefont {Saleh}},\
  }\href {\doibase 10.1088/0954-8998/1/2/006} {\bibfield  {journal} {\bibinfo
  {journal} {Quantum Optics Journal of the European Optical Society Part B}\
  }\textbf {\bibinfo {volume} {1}},\ \bibinfo {pages} {153} (\bibinfo {year}
  {1989})}\BibitemShut {NoStop}%
\bibitem [{\citenamefont {TUCCI}(1991)}]{doi:10.1142/S021797929100033X}%
  \BibitemOpen
  \bibfield  {author} {\bibinfo {author} {\bibfnamefont {R.~R.}\ \bibnamefont
  {TUCCI}},\ }\href {\doibase 10.1142/S021797929100033X} {\bibfield  {journal}
  {\bibinfo  {journal} {International Journal of Modern Physics B}\ }\textbf
  {\bibinfo {volume} {05}},\ \bibinfo {pages} {545} (\bibinfo {year} {1991})},\
  \Eprint {http://arxiv.org/abs/https://doi.org/10.1142/S021797929100033X}
  {https://doi.org/10.1142/S021797929100033X} \BibitemShut {NoStop}%
\bibitem [{\citenamefont {Agarwalla}\ \emph
  {et~al.}(2017{\natexlab{b}})\citenamefont {Agarwalla}, \citenamefont
  {Jiang},\ and\ \citenamefont {Segal}}]{PhysRevB.96.104304}%
  \BibitemOpen
  \bibfield  {author} {\bibinfo {author} {\bibfnamefont {B.~K.}\ \bibnamefont
  {Agarwalla}}, \bibinfo {author} {\bibfnamefont {J.-H.}\ \bibnamefont
  {Jiang}}, \ and\ \bibinfo {author} {\bibfnamefont {D.}~\bibnamefont
  {Segal}},\ }\href {\doibase 10.1103/PhysRevB.96.104304} {\bibfield  {journal}
  {\bibinfo  {journal} {Phys. Rev. B}\ }\textbf {\bibinfo {volume} {96}},\
  \bibinfo {pages} {104304} (\bibinfo {year} {2017}{\natexlab{b}})}\BibitemShut
  {NoStop}%
\bibitem [{\citenamefont {Newman}\ \emph {et~al.}(2017)\citenamefont {Newman},
  \citenamefont {Mintert},\ and\ \citenamefont {Nazir}}]{PhysRevE.95.032139}%
  \BibitemOpen
  \bibfield  {author} {\bibinfo {author} {\bibfnamefont {D.}~\bibnamefont
  {Newman}}, \bibinfo {author} {\bibfnamefont {F.}~\bibnamefont {Mintert}}, \
  and\ \bibinfo {author} {\bibfnamefont {A.}~\bibnamefont {Nazir}},\ }\href
  {\doibase 10.1103/PhysRevE.95.032139} {\bibfield  {journal} {\bibinfo
  {journal} {Phys. Rev. E}\ }\textbf {\bibinfo {volume} {95}},\ \bibinfo
  {pages} {032139} (\bibinfo {year} {2017})}\BibitemShut {NoStop}%
\bibitem [{\citenamefont {Curzon}\ and\ \citenamefont
  {Ahlborn}(1975)}]{curzon1975efficiency}%
  \BibitemOpen
  \bibfield  {author} {\bibinfo {author} {\bibfnamefont {F.~L.}\ \bibnamefont
  {Curzon}}\ and\ \bibinfo {author} {\bibfnamefont {B.}~\bibnamefont
  {Ahlborn}},\ }\href@noop {} {\bibfield  {journal} {\bibinfo  {journal}
  {American Journal of Physics}\ }\textbf {\bibinfo {volume} {43}},\ \bibinfo
  {pages} {22} (\bibinfo {year} {1975})}\BibitemShut {NoStop}%
\bibitem [{\citenamefont {Van~den Broeck}(2005)}]{PhysRevLett.95.190602}%
  \BibitemOpen
  \bibfield  {author} {\bibinfo {author} {\bibfnamefont {C.}~\bibnamefont
  {Van~den Broeck}},\ }\href {\doibase 10.1103/PhysRevLett.95.190602}
  {\bibfield  {journal} {\bibinfo  {journal} {Phys. Rev. Lett.}\ }\textbf
  {\bibinfo {volume} {95}},\ \bibinfo {pages} {190602} (\bibinfo {year}
  {2005})}\BibitemShut {NoStop}%
\bibitem [{\citenamefont {Lee}\ \emph {et~al.}(2018)\citenamefont {Lee},
  \citenamefont {Um},\ and\ \citenamefont {Park}}]{PhysRevE.98.052137}%
  \BibitemOpen
  \bibfield  {author} {\bibinfo {author} {\bibfnamefont {S.~H.}\ \bibnamefont
  {Lee}}, \bibinfo {author} {\bibfnamefont {J.}~\bibnamefont {Um}}, \ and\
  \bibinfo {author} {\bibfnamefont {H.}~\bibnamefont {Park}},\ }\href {\doibase
  10.1103/PhysRevE.98.052137} {\bibfield  {journal} {\bibinfo  {journal} {Phys.
  Rev. E}\ }\textbf {\bibinfo {volume} {98}},\ \bibinfo {pages} {052137}
  (\bibinfo {year} {2018})}\BibitemShut {NoStop}%
\bibitem [{\citenamefont {Ye}\ and\ \citenamefont
  {Holubec}(2021)}]{PhysRevE.103.052125}%
  \BibitemOpen
  \bibfield  {author} {\bibinfo {author} {\bibfnamefont {Z.}~\bibnamefont
  {Ye}}\ and\ \bibinfo {author} {\bibfnamefont {V.}~\bibnamefont {Holubec}},\
  }\href {\doibase 10.1103/PhysRevE.103.052125} {\bibfield  {journal} {\bibinfo
   {journal} {Phys. Rev. E}\ }\textbf {\bibinfo {volume} {103}},\ \bibinfo
  {pages} {052125} (\bibinfo {year} {2021})}\BibitemShut {NoStop}%
\bibitem [{\citenamefont {Abebe}\ \emph {et~al.}(2021)\citenamefont {Abebe},
  \citenamefont {Jobir}, \citenamefont {Gashu},\ and\ \citenamefont
  {Mosisa}}]{abebe2021interaction}%
  \BibitemOpen
  \bibfield  {author} {\bibinfo {author} {\bibfnamefont {T.}~\bibnamefont
  {Abebe}}, \bibinfo {author} {\bibfnamefont {D.}~\bibnamefont {Jobir}},
  \bibinfo {author} {\bibfnamefont {C.}~\bibnamefont {Gashu}}, \ and\ \bibinfo
  {author} {\bibfnamefont {E.}~\bibnamefont {Mosisa}},\ }\href@noop {}
  {\bibfield  {journal} {\bibinfo  {journal} {Advances in Mathematical
  Physics}\ }\textbf {\bibinfo {volume} {2021}} (\bibinfo {year}
  {2021})}\BibitemShut {NoStop}%
\bibitem [{\citenamefont {Li}\ \emph {et~al.}(2017)\citenamefont {Li} \emph
  {et~al.}}]{li2017production}%
  \BibitemOpen
  \bibfield  {author} {\bibinfo {author} {\bibfnamefont {S.-W.}\ \bibnamefont
  {Li}} \emph {et~al.},\ }\href@noop {} {\bibfield  {journal} {\bibinfo
  {journal} {Physical Review E}\ }\textbf {\bibinfo {volume} {96}},\ \bibinfo
  {pages} {012139} (\bibinfo {year} {2017})}\BibitemShut {NoStop}%
\bibitem [{\citenamefont {Sarmah}\ \emph {et~al.}(2022)\citenamefont {Sarmah},
  \citenamefont {Bansal},\ and\ \citenamefont
  {Goswami}}]{sarmah2022nonequilibrium}%
  \BibitemOpen
  \bibfield  {author} {\bibinfo {author} {\bibfnamefont {M.~J.}\ \bibnamefont
  {Sarmah}}, \bibinfo {author} {\bibfnamefont {A.}~\bibnamefont {Bansal}}, \
  and\ \bibinfo {author} {\bibfnamefont {H.~P.}\ \bibnamefont {Goswami}},\
  }\href@noop {} {\bibfield  {journal} {\bibinfo  {journal} {arXiv preprint
  arXiv:2206.07606}\ } (\bibinfo {year} {2022})}\BibitemShut {NoStop}%
\bibitem [{\citenamefont {Harbola}\ \emph {et~al.}(2012)\citenamefont
  {Harbola}, \citenamefont {Rahav},\ and\ \citenamefont {Mukamel}}]{UHeplQHE}%
  \BibitemOpen
  \bibfield  {author} {\bibinfo {author} {\bibfnamefont {U.}~\bibnamefont
  {Harbola}}, \bibinfo {author} {\bibfnamefont {S.}~\bibnamefont {Rahav}}, \
  and\ \bibinfo {author} {\bibfnamefont {S.}~\bibnamefont {Mukamel}},\ }\href
  {http://stacks.iop.org/0295-5075/99/i=5/a=50005} {\bibfield  {journal}
  {\bibinfo  {journal} {EPL (Europhysics Letters)}\ }\textbf {\bibinfo {volume}
  {99}},\ \bibinfo {pages} {50005} (\bibinfo {year} {2012})}\BibitemShut
  {NoStop}%
\bibitem [{\citenamefont {Bouton}\ \emph {et~al.}(2021)\citenamefont {Bouton},
  \citenamefont {Nettersheim}, \citenamefont {Burgardt}, \citenamefont {Adam},
  \citenamefont {Lutz},\ and\ \citenamefont {Widera}}]{bouton2021quantum}%
  \BibitemOpen
  \bibfield  {author} {\bibinfo {author} {\bibfnamefont {Q.}~\bibnamefont
  {Bouton}}, \bibinfo {author} {\bibfnamefont {J.}~\bibnamefont {Nettersheim}},
  \bibinfo {author} {\bibfnamefont {S.}~\bibnamefont {Burgardt}}, \bibinfo
  {author} {\bibfnamefont {D.}~\bibnamefont {Adam}}, \bibinfo {author}
  {\bibfnamefont {E.}~\bibnamefont {Lutz}}, \ and\ \bibinfo {author}
  {\bibfnamefont {A.}~\bibnamefont {Widera}},\ }\href@noop {} {\bibfield
  {journal} {\bibinfo  {journal} {Nature Communications}\ }\textbf {\bibinfo
  {volume} {12}},\ \bibinfo {pages} {2063} (\bibinfo {year}
  {2021})}\BibitemShut {NoStop}%
\bibitem [{\citenamefont {Yadalam}\ \emph {et~al.}(2022)\citenamefont
  {Yadalam}, \citenamefont {Agarwalla},\ and\ \citenamefont
  {Harbola}}]{PhysRevA.105.062219}%
  \BibitemOpen
  \bibfield  {author} {\bibinfo {author} {\bibfnamefont {H.~K.}\ \bibnamefont
  {Yadalam}}, \bibinfo {author} {\bibfnamefont {B.~K.}\ \bibnamefont
  {Agarwalla}}, \ and\ \bibinfo {author} {\bibfnamefont {U.}~\bibnamefont
  {Harbola}},\ }\href {\doibase 10.1103/PhysRevA.105.062219} {\bibfield
  {journal} {\bibinfo  {journal} {Phys. Rev. A}\ }\textbf {\bibinfo {volume}
  {105}},\ \bibinfo {pages} {062219} (\bibinfo {year} {2022})}\BibitemShut
  {NoStop}%
\bibitem [{\citenamefont {Dodonov}(2002)}]{dodonov2002nonclassical}%
  \BibitemOpen
  \bibfield  {author} {\bibinfo {author} {\bibfnamefont {V.}~\bibnamefont
  {Dodonov}},\ }\href@noop {} {\bibfield  {journal} {\bibinfo  {journal}
  {Journal of Optics B: Quantum and Semiclassical Optics}\ }\textbf {\bibinfo
  {volume} {4}},\ \bibinfo {pages} {R1} (\bibinfo {year} {2002})}\BibitemShut
  {NoStop}%
\bibitem [{\citenamefont {Giri}\ and\ \citenamefont
  {Goswami}(2019)}]{PhysRevE.99.022104}%
  \BibitemOpen
  \bibfield  {author} {\bibinfo {author} {\bibfnamefont {S.~K.}\ \bibnamefont
  {Giri}}\ and\ \bibinfo {author} {\bibfnamefont {H.~P.}\ \bibnamefont
  {Goswami}},\ }\href {\doibase 10.1103/PhysRevE.99.022104} {\bibfield
  {journal} {\bibinfo  {journal} {Phys. Rev. E}\ }\textbf {\bibinfo {volume}
  {99}},\ \bibinfo {pages} {022104} (\bibinfo {year} {2019})}\BibitemShut
  {NoStop}%
\bibitem [{\citenamefont {Esposito}\ \emph {et~al.}(2009)\citenamefont
  {Esposito}, \citenamefont {Lindenberg},\ and\ \citenamefont {Van~den
  Broeck}}]{PhysRevLett.102.130602}%
  \BibitemOpen
  \bibfield  {author} {\bibinfo {author} {\bibfnamefont {M.}~\bibnamefont
  {Esposito}}, \bibinfo {author} {\bibfnamefont {K.}~\bibnamefont
  {Lindenberg}}, \ and\ \bibinfo {author} {\bibfnamefont {C.}~\bibnamefont
  {Van~den Broeck}},\ }\href {\doibase 10.1103/PhysRevLett.102.130602}
  {\bibfield  {journal} {\bibinfo  {journal} {Phys. Rev. Lett.}\ }\textbf
  {\bibinfo {volume} {102}},\ \bibinfo {pages} {130602} (\bibinfo {year}
  {2009})}\BibitemShut {NoStop}%
\bibitem [{\citenamefont {Esposito}\ \emph {et~al.}(2010)\citenamefont
  {Esposito}, \citenamefont {Kawai}, \citenamefont {Lindenberg},\ and\
  \citenamefont {Van~den Broeck}}]{PhysRevLett.105.150603}%
  \BibitemOpen
  \bibfield  {author} {\bibinfo {author} {\bibfnamefont {M.}~\bibnamefont
  {Esposito}}, \bibinfo {author} {\bibfnamefont {R.}~\bibnamefont {Kawai}},
  \bibinfo {author} {\bibfnamefont {K.}~\bibnamefont {Lindenberg}}, \ and\
  \bibinfo {author} {\bibfnamefont {C.}~\bibnamefont {Van~den Broeck}},\ }\href
  {\doibase 10.1103/PhysRevLett.105.150603} {\bibfield  {journal} {\bibinfo
  {journal} {Phys. Rev. Lett.}\ }\textbf {\bibinfo {volume} {105}},\ \bibinfo
  {pages} {150603} (\bibinfo {year} {2010})}\BibitemShut {NoStop}%
\bibitem [{\citenamefont {Giri}\ and\ \citenamefont
  {Goswami}(2022)}]{PhysRevE.106.024131}%
  \BibitemOpen
  \bibfield  {author} {\bibinfo {author} {\bibfnamefont {S.~K.}\ \bibnamefont
  {Giri}}\ and\ \bibinfo {author} {\bibfnamefont {H.~P.}\ \bibnamefont
  {Goswami}},\ }\href {\doibase 10.1103/PhysRevE.106.024131} {\bibfield
  {journal} {\bibinfo  {journal} {Phys. Rev. E}\ }\textbf {\bibinfo {volume}
  {106}},\ \bibinfo {pages} {024131} (\bibinfo {year} {2022})}\BibitemShut
  {NoStop}%
\bibitem [{\citenamefont {Liu}\ \emph {et~al.}(2022)\citenamefont {Liu},
  \citenamefont {He},\ and\ \citenamefont {Wang}}]{liu2022optimized}%
  \BibitemOpen
  \bibfield  {author} {\bibinfo {author} {\bibfnamefont {H.}~\bibnamefont
  {Liu}}, \bibinfo {author} {\bibfnamefont {J.}~\bibnamefont {He}}, \ and\
  \bibinfo {author} {\bibfnamefont {J.}~\bibnamefont {Wang}},\ }\href@noop {}
  {\bibfield  {journal} {\bibinfo  {journal} {Journal of Applied Physics}\
  }\textbf {\bibinfo {volume} {131}},\ \bibinfo {pages} {214303} (\bibinfo
  {year} {2022})}\BibitemShut {NoStop}%
\bibitem [{\citenamefont {Dechant}\ \emph {et~al.}(2017)\citenamefont
  {Dechant}, \citenamefont {Kiesel},\ and\ \citenamefont
  {Lutz}}]{dechant2017underdamped}%
  \BibitemOpen
  \bibfield  {author} {\bibinfo {author} {\bibfnamefont {A.}~\bibnamefont
  {Dechant}}, \bibinfo {author} {\bibfnamefont {N.}~\bibnamefont {Kiesel}}, \
  and\ \bibinfo {author} {\bibfnamefont {E.}~\bibnamefont {Lutz}},\ }\href@noop
  {} {\bibfield  {journal} {\bibinfo  {journal} {EPL (Europhysics Letters)}\
  }\textbf {\bibinfo {volume} {119}},\ \bibinfo {pages} {50003} (\bibinfo
  {year} {2017})}\BibitemShut {NoStop}%
\bibitem [{\citenamefont {Iyyappan}\ and\ \citenamefont
  {Johal}(2020)}]{iyyappan2020efficiency}%
  \BibitemOpen
  \bibfield  {author} {\bibinfo {author} {\bibfnamefont {I.}~\bibnamefont
  {Iyyappan}}\ and\ \bibinfo {author} {\bibfnamefont {R.~S.}\ \bibnamefont
  {Johal}},\ }\href@noop {} {\bibfield  {journal} {\bibinfo  {journal} {EPL
  (Europhysics Letters)}\ }\textbf {\bibinfo {volume} {128}},\ \bibinfo {pages}
  {50004} (\bibinfo {year} {2020})}\BibitemShut {NoStop}%
\end{thebibliography}%

\end{document}